\pdfoutput=1

\documentclass[aps,preprintnumbers,amsmath,amssymb,twocolumn, tightenlines,superscriptaddress]{revtex4}

\usepackage{graphicx}
\usepackage{epsfig}

\newcommand{\be}{\begin{eqnarray}}
\newcommand{\ee}{\end{eqnarray}}

 \newcommand{\gsim}{\mathrel{\hbox{\rlap{\lower.55ex \hbox {$\sim$}}
                   \kern-.3em \raise.4ex \hbox{$>$}}}}
\newcommand{\lsim}{\mathrel{\hbox{\rlap{\lower.55ex \hbox {$\sim$}}
                   \kern-.3em \raise.4ex \hbox{$<$}}}}

\newcommand{\ba}{\begin{eqnarray}}
\newcommand{\ea}{\end{eqnarray}}

\begin{document}


\title{The Sound Edge of the Quenching Jets}
\author{Edward Shuryak and Pilar Staig}
\address{Department of Physics and Astronomy, State University of New York,
Stony Brook, NY 11794, USA.}

\begin{abstract}
When quenching jets deposit certain amount of energy and momentum into ambient matter,
part of it propagates in the form of 
shocks/sounds.   The ``sound surface", separating disturbed and undisturbed parts of the fireball, makes what we call
the sound edge of jets. In this work we semi-analytically study its shape, in various geometries. We further argue that
since hadrons with in the kinematical range of $p_\perp\sim 2\, GeV$ originate mostly from the ``rim" of the fireball,   
near the maximum of the radial flow at the freezeout surface, only the intersection of the ``sound surface" with this ``rim" would be observable. The resulting ``jet edge" has a form of extra matter at the elliptic curve, in $\Delta \phi, \Delta \eta$ coordinates, with radius 
 $|\Delta \phi| \sim |\Delta \eta |\sim 1$.  In the case of large energy/momentum deposition $\sim 100 \, GeV$ we argue that
 the event should be considered as two sub-events, with interior of the ``sound surface" having modified radial and
 directed flow. We further argue that  in the kinematical range of $p_\perp\sim 3\, GeV$ the effect of that can be large enough
 to be seen on event-by-event basis. If so, this effect has a potential to become a valuable tool to address geometry of
 jet production and quenching. 
\end{abstract}

\maketitle
    \section{Introduction}
    The hydrodynamical description of a fireball of hadronic matter created
 in heavy ion collisions is very successful, and by now it needs no introduction.
 In the last few years it has been supplemented by extensive discussion of the higher
 angular  harmonics of the flow, also successfully described by (viscous) hydrodynamics, see e.g.
 our work \cite{Staig:2011wj}.  Last year we have seen data on very peripheral AA
 collisions and the highest-multiplicity pp and pA collisions, in which the size of the fireball is quite small,
 and yet one finds the radial, $v_2$ and $v_3$ flows are in agreement with
 the ``acoustic systematics"  \cite{Lacey:2013is}  based on  viscous hydrodynamics.

  The main source of these higher harmonics of hydro expansion -- or    
   sounds as one can refer to them -- are  perturbations of the initial energy/entropy deposition of matter 
   in the transverse plane.    
      Emission of sounds $during$ the hydrodynamical evolution has not been yet been experimentally observed, although
   its existence and magnitude can be predicted from the fluctuation-dissipation theorem \cite{Kapusta:2011gt}.
   A specific model of such ``late-time" emission of sounds, by the
    near-$T_c$ collapse of the QGP clusters, is developed in our recent work \cite{Shuryak:2013uaa}.
    
     This paper is devoted to sounds emitted by another obvious perturbation: the quenching jets.  The idea, that once the
energy is deposited into the medium by a jet will be resulting in
sound perturbations in the shape of  the Mach cone, has been
proposed\footnote{ Similar idea discussed in 1970's for non-relativistic
nuclear collisions did not work, because the mean free path in nuclear matter is large, $1-2\, fm$.
Effective mean free path in sQGP under discussion now is about an order of magnitude smaller, and hydrodynamics
works quite well.
}
 in Refs \cite{Casalderrey-Solana:2004qm,Horst}.

    From theoretical perspective  the fate of the deposited energy by a local source has been 
addressed in the framework of AdS/CFT, the only tool we have for first-principle dynamical treatment
 of strongly coupled QGP. As shown by Chesler and Yaffe
 \cite{Chesler:2007an}, and Gubser et al
\cite{Gubser:2009sn}, the stress tensor solution obtained by the solution of the
first principle Einstein equation
and the so called holographic imaging,  was found to be in remarkably good agreement with
the hydrodynamical solution obtained in
\cite{Casalderrey-Solana:2004qm}. 

While the theoretical developments of the effects are quite solid, its phenomenological implementation 
turned out to be rather controversial,  as it was first confused with the effect of the initial 
state fluctuations mentioned above. When PHENIX and STAR collaboration had studied the two-particle correlation functions
at $p_\perp \sim 3 \,GeV$, they had found two peaks at $\Delta \phi=\pm 2\, rad $ instead of the  associated jet peak expected at $\Delta \phi=\pi$. Those peaks has been interpreted 
 \cite{Casalderrey-Solana:2004qm,Horst}  as manifestation of the Mach cone.
 Since the
time-averaged sound velocity over the QGP, mixed and hadronic
phases is
$
<c_{s}>\,\approx\,0.4\,,
$
the expected Mach cone angle
\be %
\theta_{M} = arccos \left({<c_{s}> \over v_{jet} } \right) \approx
1.1\,rad  \label{eqn_Mach}
\ee %
from the associated jet  indeed matches their angular positions; $\pi\pm \theta_{M}\approx \pm 2.0$.
 However, as it turned out, that was a mere coincidence. Presumed dominance of jets for
  secondaries with $p_\perp \sim 3 \, GeV$ was in fact misleading, and in reality this kinematic window is instead
dominated  by the tail of the hydro flow. The double peaks at 
$\Delta \phi=\pm 2\, rad $ are due to large triangular flow and related to  the ``sound horizon", as detailed e.g. in \cite{Staig:2010pn}.

 As the $p_\perp$ of the trigger hadron is increased further, the expected dominance of jet-related effects does appear.
   And  
      once the fraction of the energy/momentum of the jet
 is deposited into the medium locally, hydrodynamics  requires the appearance of
shock/sound perturbations. It is as inevitable as thunder after the strike of the lightning. 

    Let us now briefly refer to the previous studies of the problem. The original Mach cone solution 
    \cite{Casalderrey-Solana:2004qm} was demonstrated for the simplest case of infinite homogeneous matter and constant 
 energy deposition $dE/dx$. Effects of the fireball explosion on the Mach cone 
 has been discussed by Satarov, Stoecker and Mishustin, \cite{Horst}, as well as by 
 Betz,Rau and Stoecker \cite{Betz:2007ie}, see also their subsequent works.
 Khachatryan and Shuryak 
 \cite{Khachatryan:2011px} studied the problem using  the ``geometric acoustics" approximation.
   
The present paper carries those studies further. We first point out a very significant simplification: 
 due to very strong radial flow,  the secondaries with $p_\perp \sim 2 \, GeV$
  are emitted from only a small fraction of the freezeout surface, near the so called  ``rim of the fireball".
 Therefore, (i) if one selects the associate particles in this kinematical window,
   only the $overlap$ of this rim with the sound  surface are observable.
 This simplification allows us to predict the relation between the  event geometry and
the  shape of its contribution to the two-particle correlation function, as we detail below. 

The second and third elements are that we not only include distortions of  the ``sound surface" due to
the background flow, but also include (ii) the realistic viscosity, as well as (iii) an inhomogeneous energy deposition by the jet. These two allow for a more realistic estimate of the perturbation amplitude.

The perturbative BDMPS  theory \cite{Baier:1994bd} predicts the jet quenching to be dependent on the time since jet origination: it grows proportionally to it,
$dE/dx\sim x$, while the strong coupling AdS/CFT approach  suggests even stronger dependence $dE/dx\sim x^2$.
Some data indicate more complicated dependence of $dE/dx$ on the matter temperature, with a peak at $T=T_c$ 
 \cite{Liao:2008dk}: for recent phenomenological updates on ``jet tomography" see \cite{Zhang:2012ha,Betz:2013caa}.
(We will use below only the simplest of those, the BDMPS one.)
The combined effect of viscosity and inhomogeneous deposition  substantionally change the amplitude of the perturbation, placing more emphasis to the later stages of the process. They  
weaken the Mach cone  and enhance the role of  the last deposition point.  

  The outline of the paper is as follows. In section \ref{sec_geometry} we discuss general geometrical features of the sound surface and outline some qualitative effects.
     We then identify  the best kinematical range for the associate particle: as we want as large contrast 
  of the transfer of information from the collective motion at the freezeout to the detector, it should be at the {\em upper edge}
     of the hydro-dominated region, so the best choice is $p^{A}_\perp=2-3 \, GeV$.

   The central part of the paper is the  section \ref{sec_examples} in which we provide a number of examples in which the hydrodynamical perturbations of the flow are
   calculated. The method used is based on relatively simple analytic solution for central collisions known as Gubser flow \cite{Gubser:2010ze}. The linearized equations 
for perturbations on top of it depend on all 4 variables, but using cleverly designed comoving coordinates \cite{Gubser:2010ui} one finds that 
dependence of the solution on all four coordinates can be written as separable set of functions. In our previous work \cite{Staig:2011wj}
a complete Green function for perturbations  has been already evaluated, and the solutions reported here are basically a convolution of
this Green function with the jet energy deposition $dE/dx$ along the jet path.
 
Section \ref{sec_pheno} is devoted to discussion of phenomenological information, coming from RHIC and LHC experiments. 
We start it from the ``low energy" end of the jet spectrum, in which case one can use dihadron correlation function with a trigger hadron $p^T_\perp\sim 10 \, GeV$.
As in the rest of the paper, the best kinematical window for the associate hadron is $p^A_\perp\sim 2 \, GeV$.
One may perhaps benefit from using the identified protons/antiprotons, rather than all charged particles.

  The last part of   section \ref{sec_pheno}   is related with the
    discovery at LHC high energy dijets with large asymmetry, with the  energy/momentum deposition
  as large as $\sim 100 \, GeV$. As we will see below, it is
    large enough to affect part of the underlying event, perhaps to the extent visible on event-by-event basis.

    \section{Geometry} \label{sec_geometry}
 \subsection{The sound surface}   
 
 \begin{figure}[t]
\includegraphics[width=7cm]{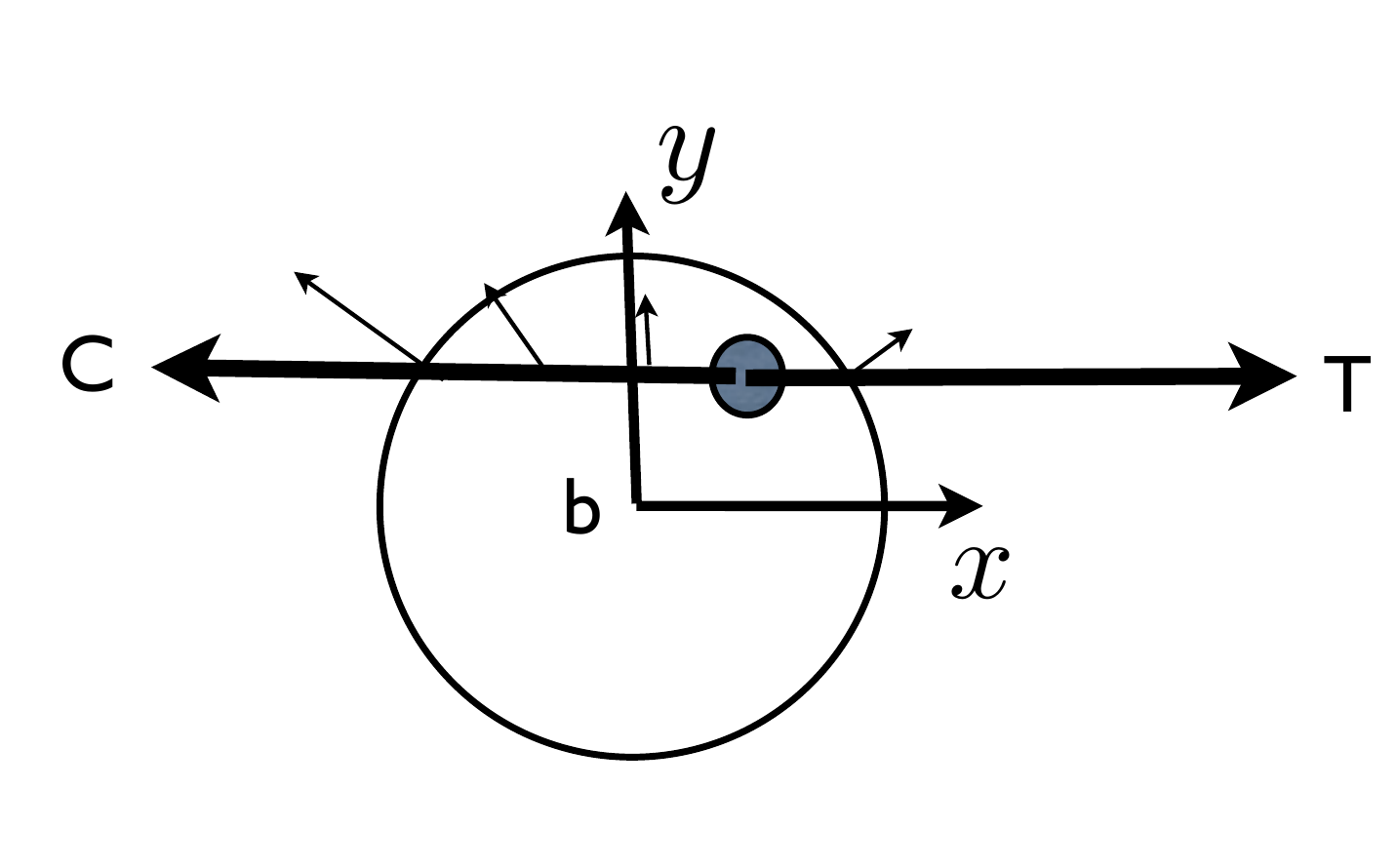}
  \caption{A sketch explaining notations: $x,y$ are coordinates   in the transverse plane.
   The trigger jet T is, by definition, emitted in $x$ direction, $\phi=0$ and the companion jet C opposite to it, $\phi=\pi$. The jet path has impact parameter $b$ in respect to the fireball center.  Thin arrows indicate direction of the radial flow, whose magnitude grows with time 
    approximately linearly.  }
  \label{fig_cone}
\end{figure}

  The notations we use are indicated in Fig.\ref{fig_cone}: the trigger jet T goes in $+x$ direction, and thus its companion jet C
  in  $-x$ direction, with the impact parameter $b$ in respect to the fireball center. 
   For simplicity, in this paper we only consider (near) central collisions, thus the circle represents an axially symmetric fireball.
   Since
trigger-bias force the companion jet to deposit much larger amount of energy, the former one has much larger chance to become visible.

\begin{figure}[t]
\includegraphics[width=7cm]{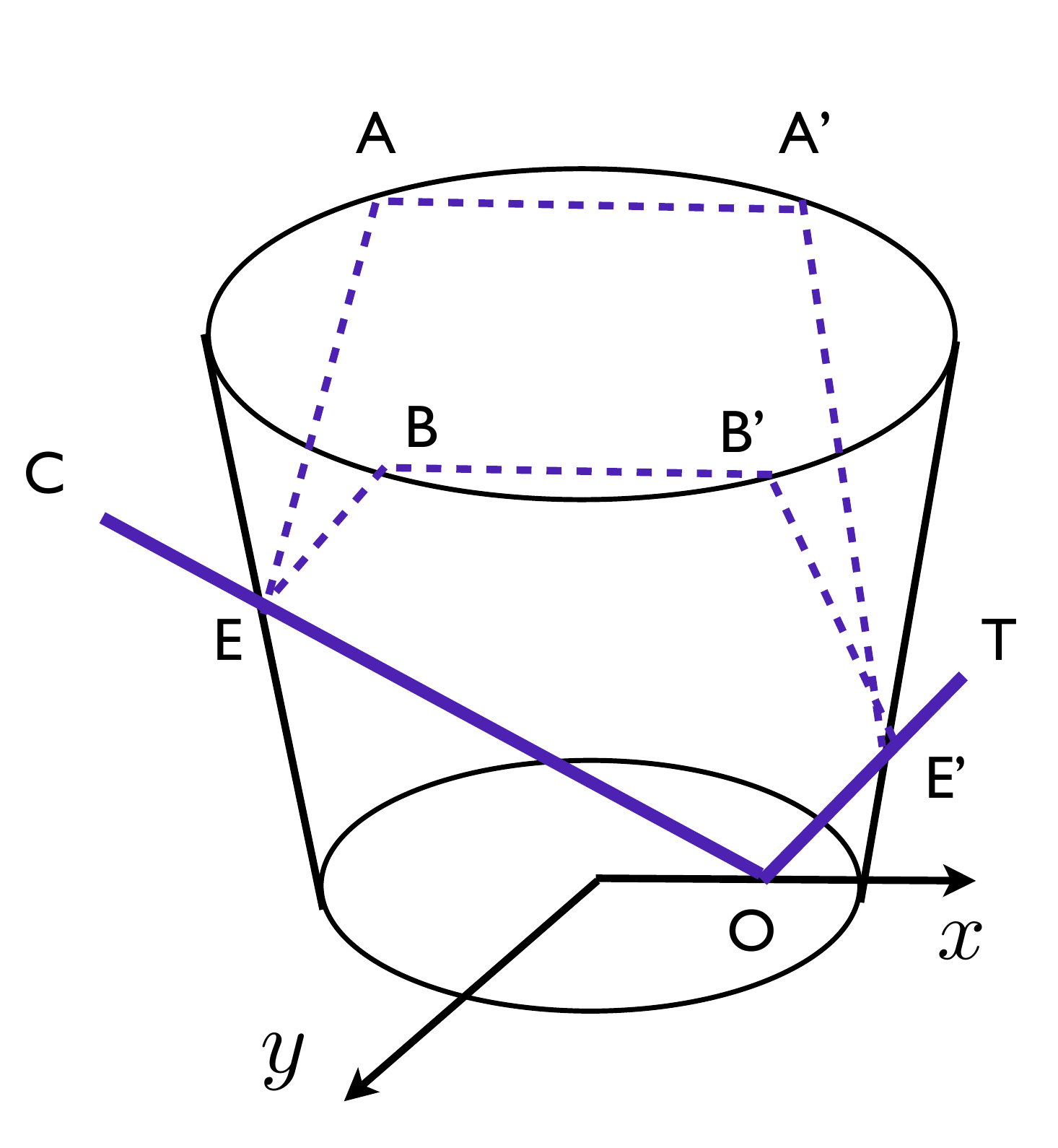}
\includegraphics[width=8cm]{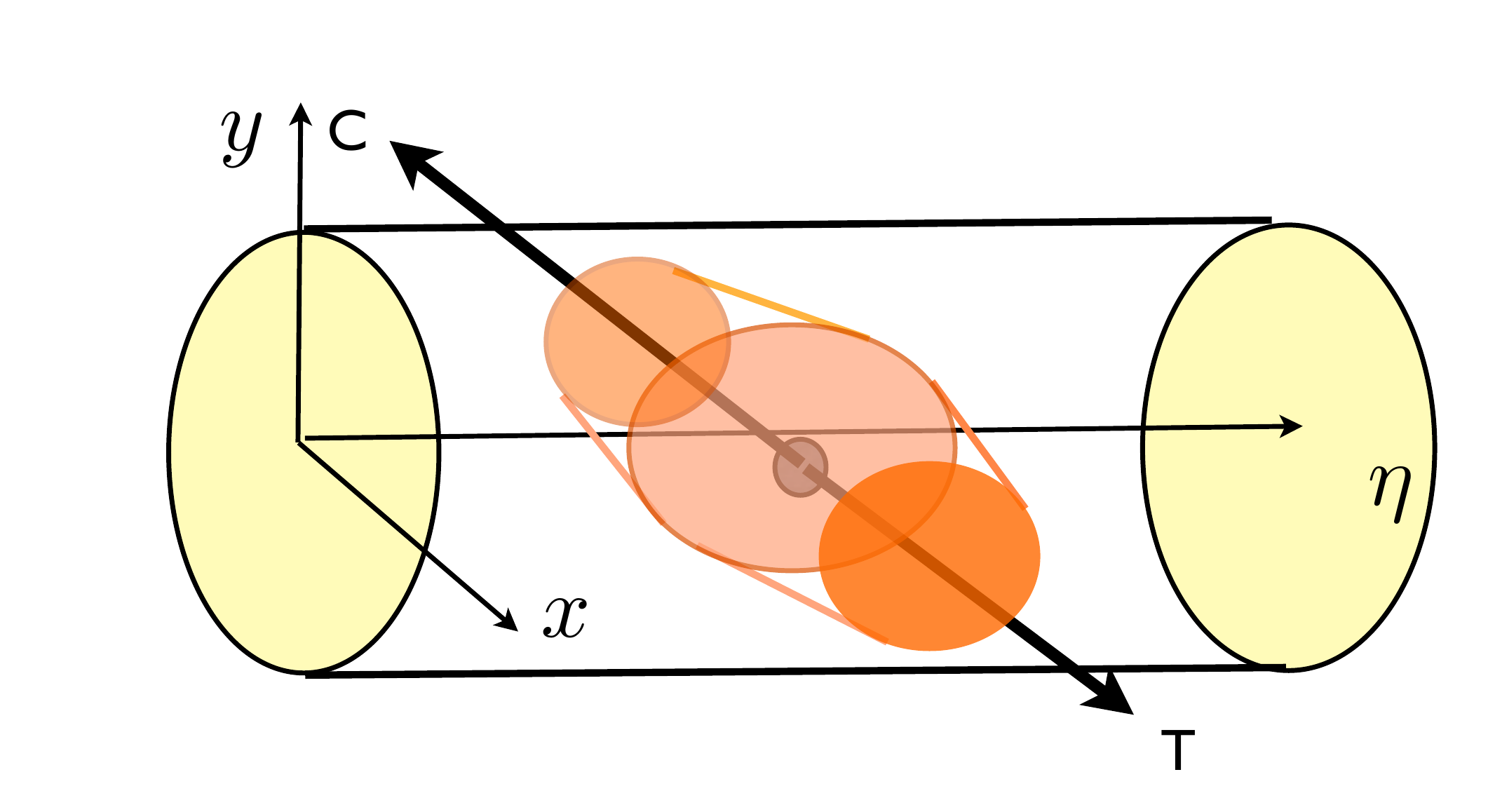}
  \caption{ (a) Schematic shape of the (2d) sound surface in the (3d) picture including the transverse coordinates $x,y$ and 
  the proper time $\tau$ (vertical direction). The lower and upper circles indicate the initial and final surfaces. The jet
  origination point is called O, the two exit points are E and E'. 
  The sound surface consists of two parts, $OEAA'E'$ and $OEBB'E'$, indicated by the dashed lines. The value $b=0$ is chosen for simplicity.\\
  (b)  A schematic view of the sound surface in coordinates  $x,y$ and (spatial) rapidity $\eta$. Trigger and companion jets are
chosen to have  the same rapidity, for simplicity. }
  \label{fig_cone2}
\end{figure}

 Schematic picture of the sound surface is depicted in Fig.\ref{fig_cone2}. Its part (a) adds to the transverse plane $x,y$
 the (longitudinal proper) time \be \tau=\sqrt{t^2-z^2} \ee which runs vertically upward. 
 The lower and upper circles thus indicate the initial and final time-like surfaces at which
 hydrodynamics starts and ends.
 The trigger and companion jets $T,C$ exit the fireball at points
 $E',E$.   The sound surface --defined as the one separating the sound-disturbed and undisturbed parts of the fireball  -- is
  indicated by the (blue) dashed lines, it consists of two parts $OEAA'E'$ and $OEBB'E'$. By hydro causality, 
  only the interior part of the fireball can absorb the energy and momentum deposited by the quenching jet.
  
The case shown is a typical one, but below the reader will see examples of other possibilities. In general, 
    those can be enumerated as follows:
  \\ 
   I. The trigger jet is assumed to leave the fireball. The companion
   jets which leave the fireball -- called  ``punched-through jets" --  can do so
by   leaving through the timeline part of the freezeout surface (case Ia), or, much more likely,  through the space-like
   part of it as shown in Fig.\ref{fig_cone2}(a) (case Ib).\\
   II.The companion jet can be stopped inside the fireball: in this case the surface is complemented by a (distorted) sphere around
   the final point.  \\
   What happens in a specific event depends on the original point and direction of the jet, as well as on its energy and, of course, quenching $dE/dx$
and  global observables, such as the collision energy and the impact parameter of the collision.  
 
 In Fig.\ref{fig_cone2}(b) we show another view of the sound surface, now at some late (freezeout) time
 as a function of transverse coordinates $x,y$ complemented by the so called space-time rapidity
 \be \eta={1\over 2} ln{t+z \over t-z} \ee In this variable the fireball looks like a long and nearly-homogeneous cylinder,
 between two ``lids" containing fragmented remnants of the original nuclei. The ``sound surface" is also tube-shaped,
with the widest section around the jet origination point O and containing two Mach cones terminating at the fireball edge.
The four points $A,A'$ and $BB'$ at fixed $\eta$ indicated in Fig.(a) are in fact pairwise connected, forming two elliptic curves around T and C jets,
which will play significant role in what follows.   

 \subsection{Kinematics and  the role of the ``fireball rim"}

   Particles observed in the detectors come from the  freezeout surface $\Sigma$ and their distribution is 
   written as the so called Cooper-Fry formula
\begin{eqnarray}
dN & = \frac{d^3p}{ (2 \pi)^3 E} p^{\mu} & \int_{\Sigma}d^3\Sigma_{\mu} 
exp(p^{\mu}u_{\mu}/T_f).
\end{eqnarray}
where $p^{\mu}$ is the 4-momentum of the particle, $d\Sigma_{\mu}$ is the vector normal to the freeze-out surface.     In most hydrodynamical applications to heavy ion collisions, the  freezeout surface is approximated by an isotherm with some
    $T=T_f\sim 100 \, MeV$.The exponent is the equilibrium distribution in Boltzmann approximation. 
    
    (Note, that the validity region of
this approximation is limited from below -- very soft pions -- in which Boltzmann exponent should be replaced by the Bose-Einstein distribution, and also from above,
by some $  p_\perp^{max}\sim 3\, GeV$ at which the viscosity-induced corrections to the distribution $\delta f$
induced by the flow gradients   become large.)

We use notations $p_z = m_\perp \sinh{y_p}$, where $y_P$ is particle rapidity and $m_\perp^2=m^2+p_\perp^2$ is the so called transverse mass
 so $dp_z = m_\perp \cosh{y_p}dy_p =Edy_p$.  
 Throughout this calculation we will work in $(\tau, r,\phi, \eta)$ coordinates, where the four-momentum is written as
\ba
p^{\mu} &=& ( m_\perp\cosh{(y_p-\eta)}, p_\perp\cos{(\phi_p-\phi)},\frac{p_\perp}{r}\sin{(\phi_p-\phi)}, \nonumber \\ 
&&  \nonumber \frac{m_\perp}{\tau}\sinh{(y_p-\eta)} ) 
\ea
where $\phi_p$ and $\phi$ refer to azimuthal angles in the momentum and position spaces.

The first general observation
is that only a fraction of the surface $\Sigma$ contributes to the production of particles
with a particular 4-momentum. Indeed, one would expect that the angular directions close to 
those of the momentum $y\approx \eta,\phi_p\approx \phi$ would contribute more to the integral.
Furthermore, this tendency should be enhanced with $p_\perp$, eventually reducing the important
integration region into a small spot 
 and allowing for the saddle-point approximation 
in the transverse plane, see the early paper \cite{Blaizot:1986bh} and more detailed discussion in our 
paper \cite{our_harmonics}.

 Indeed, let us single out one term in the exponent governing the $\phi$ integral, namely the one containing $\cos{(\phi_p-\phi)}$.
Its coefficient
 \be  A={p_\perp \over T_f} sinh(\kappa) \approx 26
 \ee
in which we have introduced the transverse rapidity of the flow $\kappa$ and use $u_r=sinh(\kappa)$ in the r.h.s. , substituted some
 typical values  $p_\perp=2.4 \, GeV$ and $T_f=.12 \, GeV$ and the maximal flow $\kappa\approx 1.1$.
Since it is in the exponent and 
$A\cos{(\phi_p-\phi)}\approx A-((\phi_p-\phi)^2 A/2$, the angular integral in $\phi$ is approximately Gaussian. 
  Ignoring for now pre-exponent, one can write those two as well known generic integrals
  \be  \int_0^{2\pi} {d\phi \over 2 \pi} exp[A cos(\phi -\phi_p)]=J_0(A)\approx exp(A){1 \over \sqrt{2\pi A}} \nonumber  \ee
  \be \int_{-\infty}^{\infty} d\eta e^{-A cosh(\eta -y_p)}=2 K_0(A)\approx \sqrt{{2\pi \over A}} exp(-A) \nonumber \ee 
  where the right expressions are asymptotics at large $A$. For  $A$ in the realistic range the
  asymptotical expressions  work reasonably well. Therefore,
the
width of the contributing ``spot" in $\phi$ integration is thus indeed  small:
\be \sqrt{<(\phi - \phi_p)^2>} = {1\over \sqrt{A}}\approx  {1 \over 5} \ee
compared to its total period $2\pi$. 
Similar conclusion follows about the spatial rapidity deviations from that of the hadron $y-\eta$.

 The consideration of the $r$ integration is a bit more involved.  The positive and negative terms in the Boltzmann exponent   to a large extent cancel each other, due to collinear relativistic motion.
   This is because the particle energy in the frame coming with a flow 
   is very different from $p_\perp$ in the lab frame: this cancellation enhances the production rate by many orders of magnitude. 
   Indeed, the thermal factor  have in the exponent the energy in the frame  comoving with the flow 
   \be B=  {p^\mu u_\mu \over T_f}\approx {p_0 \over T_f} \sqrt{{1-v_\perp \over 1+v_\perp}}= {p_0 \over T_f}e^{-\kappa} \approx 7
     \label{eqn_param}  \ee
     where, we remind, $\kappa$ is the transverse flow rapidity and $v_\perp=tanh\kappa $ is the transverse flow velocity.
     The reduction from $A$ to $B$, or an increase of the ``apparent temperature", is due to $exp(-\kappa)$, 
  the so called ``blue shift" factor. Note that it still leaves us with a numerical value of $B$
        large enough to be used as a large parameter, reducing the integral over $r$ plane to a vicinity of a  point
      $r=r^*$ at which the transverse flow has its maximum.

      \begin{figure}[t]
\begin{center}
\includegraphics[width=6 cm]{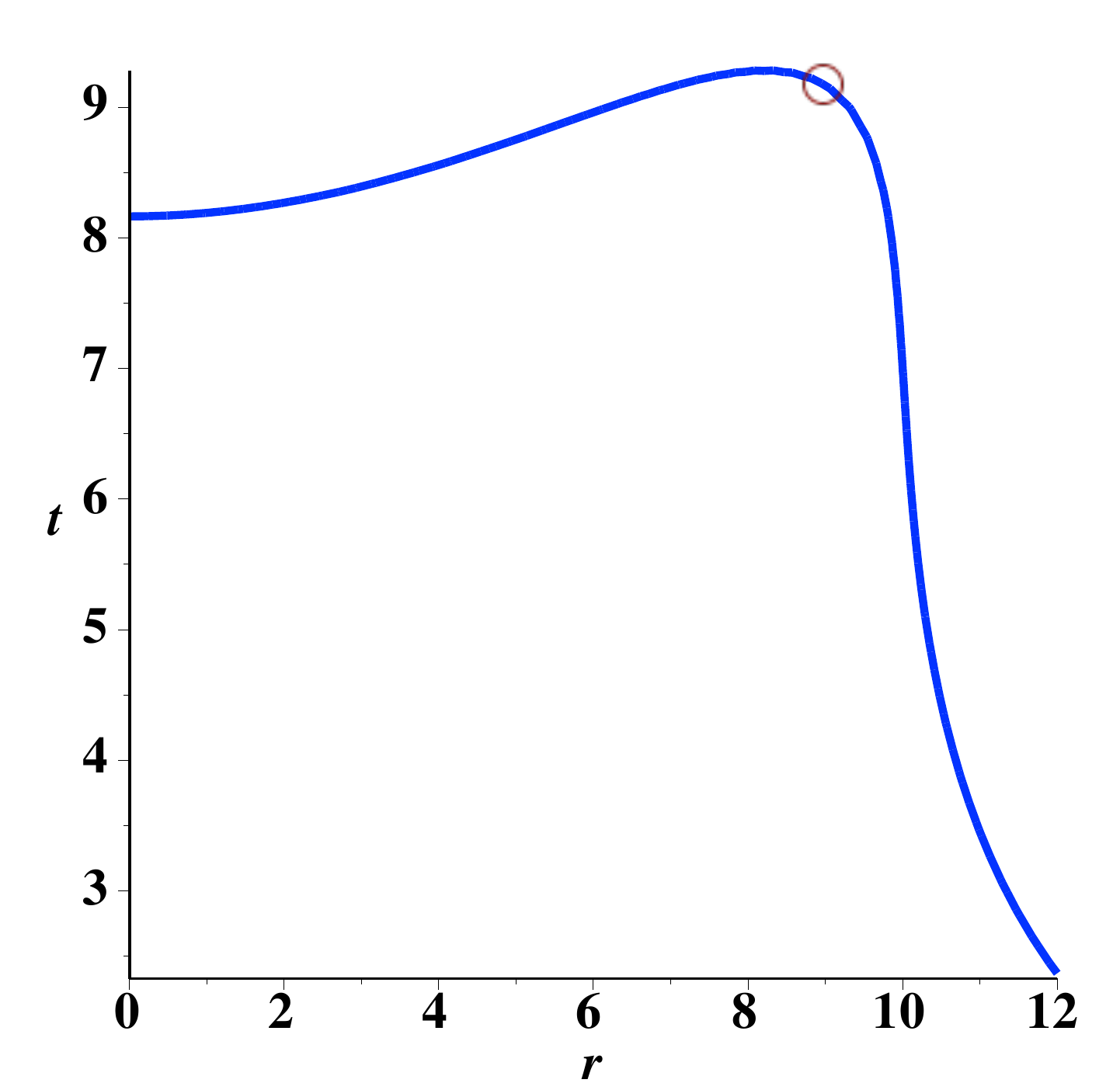}
\includegraphics[width=6 cm]{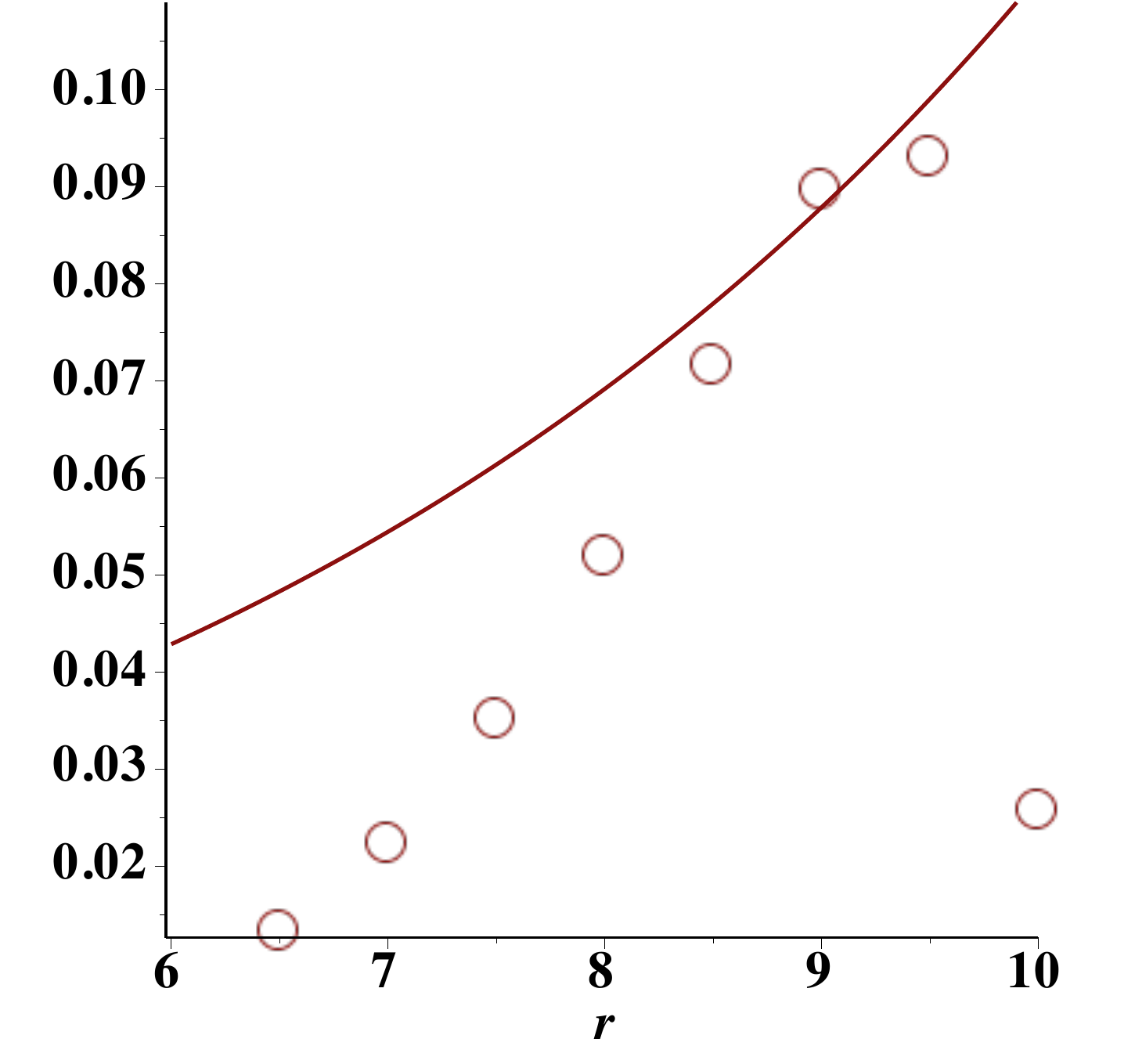}
\end{center}
\vspace{-5ex}\caption{(Color online) (a) Example of a  freezeout surface with $T_f=120 \, MeV$ surface in the $r,\tau$ plane (both in fm). The circle on the line
indicate a point in which the transverse flow reaches its  maximal  value $v_\perp=0.89$.
(b) The circles show the radial dependence of the expression (\ref{CF_radial}), the line indicates the exponential
approximation discussed in the text.
}\label{fig_freeze}
\end{figure}
   
   A typical shape of a freezeout surface is shown in Fig. \ref{fig_freeze}(a), where we also indicated the location of the maximal flow point. 
   In Fig. \ref{fig_freeze}(b)    we display the radial dependence of the main part of the Cooper-Fry integrand
   \be r *exp\left(-{p_t \over T_f}   \sqrt{{1-v_\perp \over 1+v_\perp}} \right) \label{CF_radial} \ee
  for $p_t =2.4 \, GeV,T_f=.12 \, GeV$, with the velocity taken
     at the freezeout surface $v_\perp(\tau_{FO}( r) ,r)$. We observe that indeed
    there is a sharp maximum  near the point of the maximal blue-shift, and so the integral is dominated by the ``rim".
    
    (While     the integrand appears to be falling sharply  on the right side of that point,      
 detailed studies show that at large $p_\perp$ the r.h.s. of the peak start contributing to the spectrum.
 Here we in fact find limitations of the Gubser's flow model:  it 
 has initial tail of the matter density which is power-like, and the correspondent freezeout boundary move
    {\em inward}, from $r\sim 14 \, fm$ at early time to $r \approx 9\, fm$ at late time.
     Realistic nuclear density falls exponentially, and hydro calculations show very different 
     shape of the space-like part of the freezeout surface: it moves {\em outward}, from   $r\approx 6$ to $r\approx 9.1$.
     
     More generally, the outer wall of the freezeout surface is the place of large gradients, leading to large viscous corrections.
     In current literature those gradients are included to the first or second order, but --
     to our knowledge -- convergence of those expressions at the outer wall region has not been convincingly shown.
     
Fortunately, this outer part of the surface moves rather quickly, $p_\mu\Sigma_\mu$ is small, and for most of the spectra only   few  percents of particles
come from it. As far as this fraction is considered negligible, the
     exact shape of the outer wall of the freezeout surface is unimportant. However as the $p_\perp$ grows beyond  $p_{max}\sim 3\, GeV$ or so, this region 
     becomes relevant. So, at this point, we treat it as an open problem, suggesting for now      
    to include only the time-like part of the freezeout surface. )

    In summary,
 for $p_\perp\sim 2\, GeV \gg T_f $ strong radial flow in the exponent of the Boltzmann factor reduces the Cooper-Fry
 integral  to only a small spot on the freezeout, limited in $\phi,\eta$ and in $r$ being close
to the radius $r_{rim}$ at which the radial flow is maximal. We will call location of those points ``the rim
 of the fireball",
 for axially symmetric central collisions in the setting we discuss it is a circle of radius $9.1\, fm$. 

  So far we had only considered the ``background" radial flow, without the temperature and 
  flow velocity perturbations. As will be shown by calculated examples below, the maximal perturbations induced by jets are
 located near the sound surfaces.  The conclusion following from this statements
 and preceding kinematical discussion is that the high end of the hydro spectrum is dominated by the {\em intercept}
 of the sound surface and the fireball rim. 
 
 Our main aim in the calculation below would
 thus be to provide a number of examples in which we indicate location of this intercept. 
As already outlined in the introduction above, 
  in variables $\phi,\eta$ it will have form of certain elliptic curves, we call the ``jet edges".
  
   We will not go into detailed calculation of the particle spectra, and only indicate that the $\phi_p,y_p$ spectra of associate hadrons are a  (somewhat blurred) copy of $\phi,\eta$ distributions we calculate.
The resolution (blurring) of the transition between those two sets of variables is given by
smallness of the spot size $1/\sqrt{A}$ defined above.

 \section{Sounds from jets: examples}  \label{sec_examples}
 \subsection{Perturbations of the Gubser flow}
 
     The original solution has been found by Gubser \cite{Gubser:2010ze} and the perturbation analysis
     was done  by Gubser and Yarom  \cite{Gubser:2010ui}. We do not replicate here many
     expressions from those papers: the reader interested in technical details should consult them. Some information can be also
     found in our paper \cite{Staig:2011wj} in which we basically defined  the Green function for perturbation
     from extra entropy (energy, temperature) deposited in a delta-function-like way at certain space-time point. 
     Jet perturbation is a convolution of the energy deposition function with this Green function.
 The only dimensional parameter of the model $q$  is taken below to be $1/q=4.3\, fm$, as
in the original Gubser's paper  \cite{Gubser:2010ze},  to approximate heavy nuclei $Au,Pb$
used at RHIC,LHC.
   
   The  basic coordinates used are hyperbolic pair $\tau,\eta$ for longitudinal coordinates --
already defined above --    and the polar coordinates $r,\phi$ in the transverse plane. 
However equation of motion for perturbations become separable in different --
    comoving -- coordinates        \cite{Gubser:2010ui} , substituting $\tau,r$ by
\begin{eqnarray}
\sinh{\rho} & = & -\frac{1-q^2\tau^2+q^2r^2}{2q\tau}\label{rho_coord}\\
\tan{\theta} & = &
\frac{2qr}{1+q^2\tau^2-q^2r^2}\label{theta_coord}
\end{eqnarray}   
         The dimensionless temperature (such quantities are denoted by a hat) $T\tau$
is only a function of $\rho$.      In ideal approximation (without viscosity) it is given by especially simple expression
   \be  T(r,\tau) ={\hat{T_0}  \over  f_*^{1/4}  } {1 \over \tau cosh^{2/3} (\rho)}  \ee 
  where the parameter $  
f_* = \frac{\epsilon}{T^4} \approx 11$ according to QGP thermodynamics. For the LHC conditions we had selected in 
    \cite{Staig:2011wj} the value $\hat{T_0}  =10.1$. 
    The freezeout surface we define as the isotherm $T=120\, MeV$.
    
    General solution for the linearized system of equations (we will not repeat here) can be written as a sum
   \be  {\delta T \over T}=\delta(\rho,\theta,\phi,\eta)= \sum_{l,m,k} c_{k,l,m} R_{l,k}(\rho) Y_{l,m}(\theta,\phi) e^{ik\eta}\ee  
   where $Y_{l,m}$ are the usual spherical harmonics.
   The function $R$ depending on ``comoving time" is analytically known for zero viscosity, and is numerically calculated
   in the non-zero viscosity case, from the corresponding (ordinary) differential equations.
    We typically discretize the energy deposition
   into 20 events along the jet path, calculating corresponding coefficients $c(k,l,m)$ as a sum over those events.
   We also keep $20$ values of $l$ and $m=-l..l$, as well as 20 values of discretized $k$. Those multiple sums are done via fortran program. The results will be typically shown as pictures of $\delta$ at fixed $\tau$ in the transverse plane $x,y$.

  \begin{figure}[t!]
\includegraphics[width=8cm]{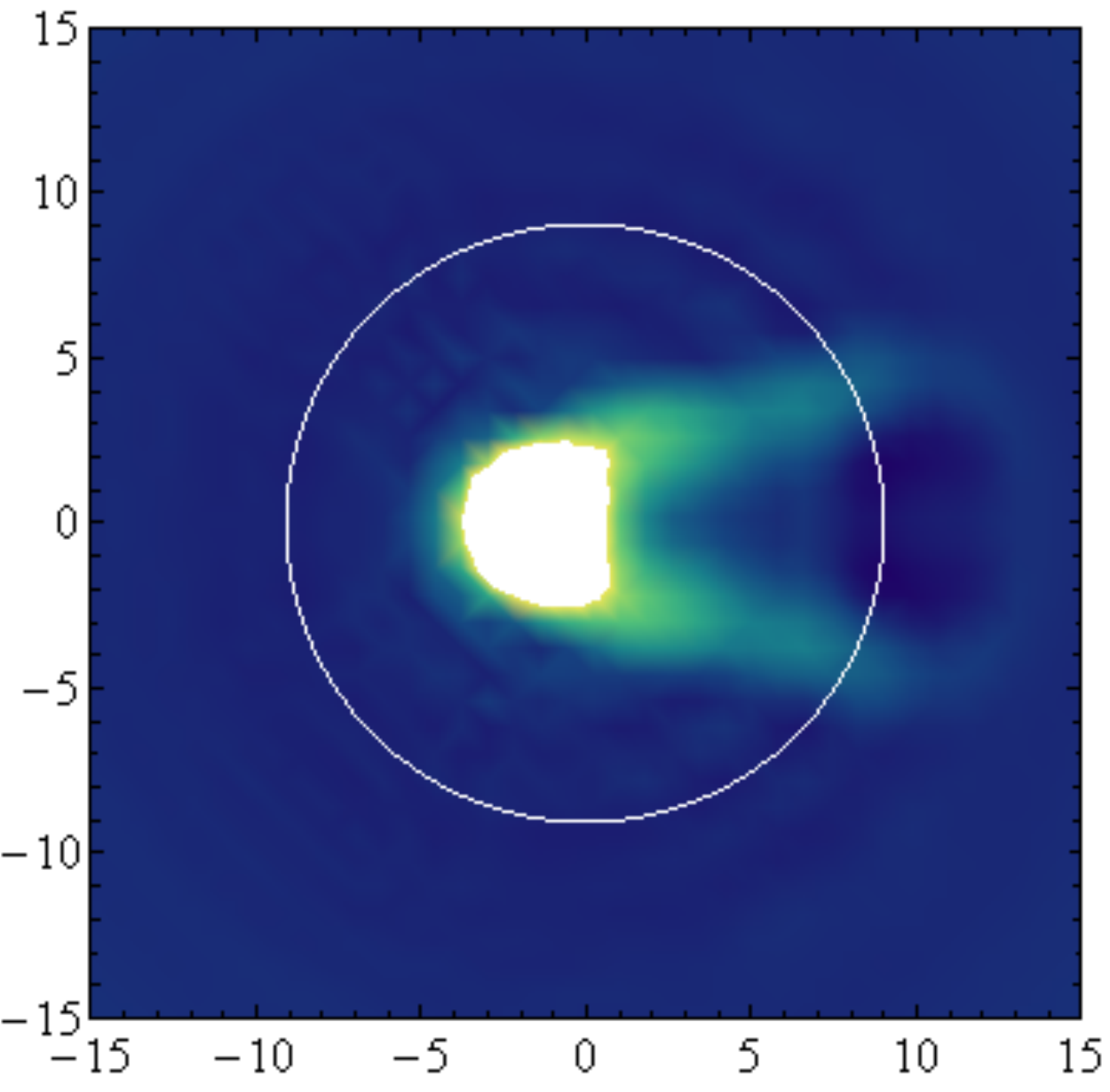}
\includegraphics[width=8cm]{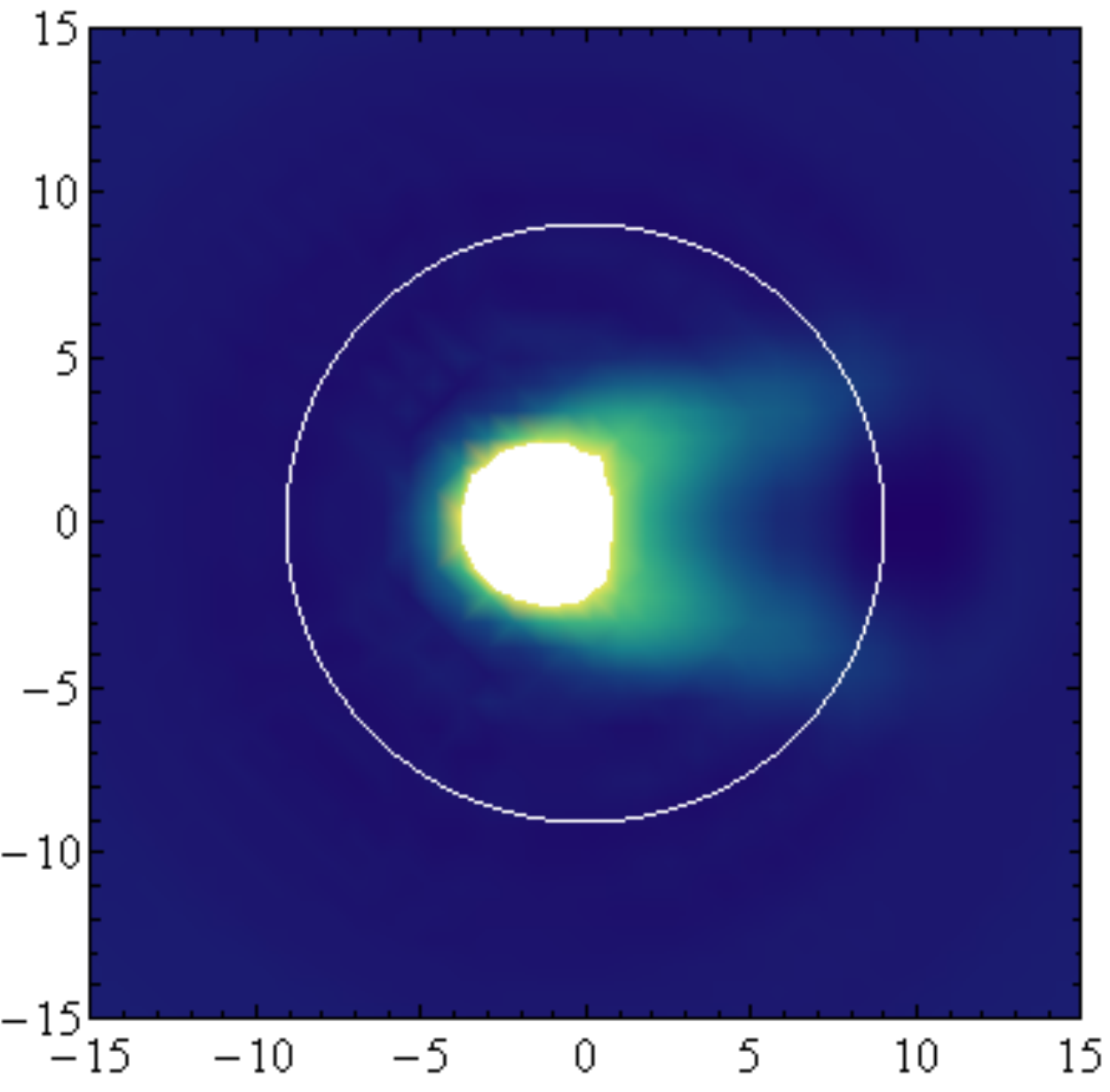}
  \caption{ Perturbation (arbitrary scale) of the temperature in the transverse plane (x,y) (in fm), induced by a jet
  generated at the point (6.1,0) and moving to the left along the diameter with the speed of light.  In the upper plot
  the energy deposition $dE/dx=const$ and viscosity is put to zero, while for the lower one 
 $dE/dx\sim x$ and viscosity-to-entropy combination being $4\pi\eta_{shear}/s=2$.   }
  \label{fig_U2}
\end{figure}
 
   \subsection{Punched-through jets}

  The calculated temperature perturbation $\delta T/T$ in the transverse plane at freezeout time is shown in several subsequent plots. For technical simplicity, we made certain approximations while producing those.
We show distribution   at  fixed proper time $\tau= 9.55\,  fm$, as an approximation to the freezeout surface.
Till section \ref{sec_eta} we
 ignore the $\eta$ variable and a sum over conjugated momentum $k$. These simplification have only minor effect
 on the plots, as we checked in few cases.

 Our first example is shown in Fig.\ref{fig_U2}, as two contour plots. They  correspond to 
 a jet moving along the diameter of the fireball and exiting via the timeline part of the freezeout surface.
 As the viscosity is switched off, and the energy deposition along the path is taken to be constant,
 one can clearly see in Fig.\ref{fig_U2}(a) the Mach cone, with its
 two arms  joined together by a circle-like perturbation  (located outside the fireball rim $r>9.1\, fm$
to be ignored).  
 It should be compared to the second one,  shown in Fig.\ref{fig_U2}(b), in which the energy deposition $dE/dx\sim x$, 
 as predicted by the BDMPS  theory \cite{Baier:1994bd}, 
 and viscosity is set to the realistic value $4\pi\eta_{shear}/s=2$ included in ``acoustic damping" formula \cite{Staig:2010pn}
 (in which we also substituted $m^2\rightarrow l(l+1)$, the value appropriate for the angular Laplacian). 
 In all subsequent plots we will use the same setting.
 
 As one can see, in this case the Mach cone is  weakened
 significantly, although 
 the triangular shape of the ``head" of the perturbation is well preserved.  
 However, as most of this perturbation does not intercept the fireball rim, we do not expect this
 case to lead to significant observable signal.

\begin{figure}[t!]
\includegraphics[width=8cm]{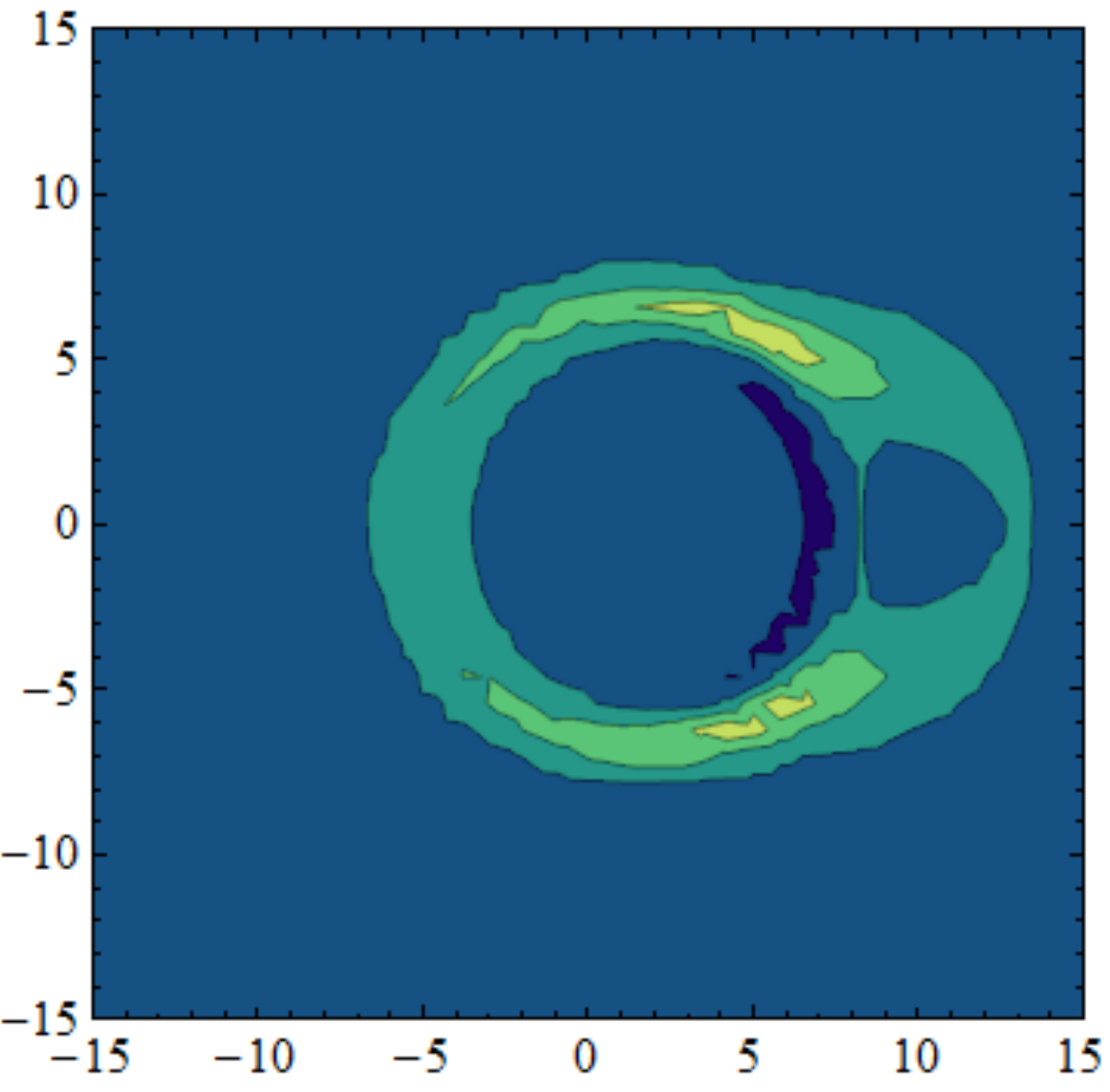}
  \caption{ Perturbation (arbitrary scale) of the temperature in the transverse plane (x,y) (in fm), induced by a jet
  generated at the point (6,0) and stopped at (0,0).   }
  \label{fig_U4}
\end{figure}

If the jet stops inside the fireball, the Mach cone gets ``rounded" by a sphere centered at the stopping point. 
  One example is 
  the jet which is originated near the fireball rim and stopped at the fireball center, as shown in Fig.\ref{fig_U4}. The
 remnants of the Mach cone are visible, now near
the  rim of the fireball. 

\begin{figure}[t]
\includegraphics[width=8cm]{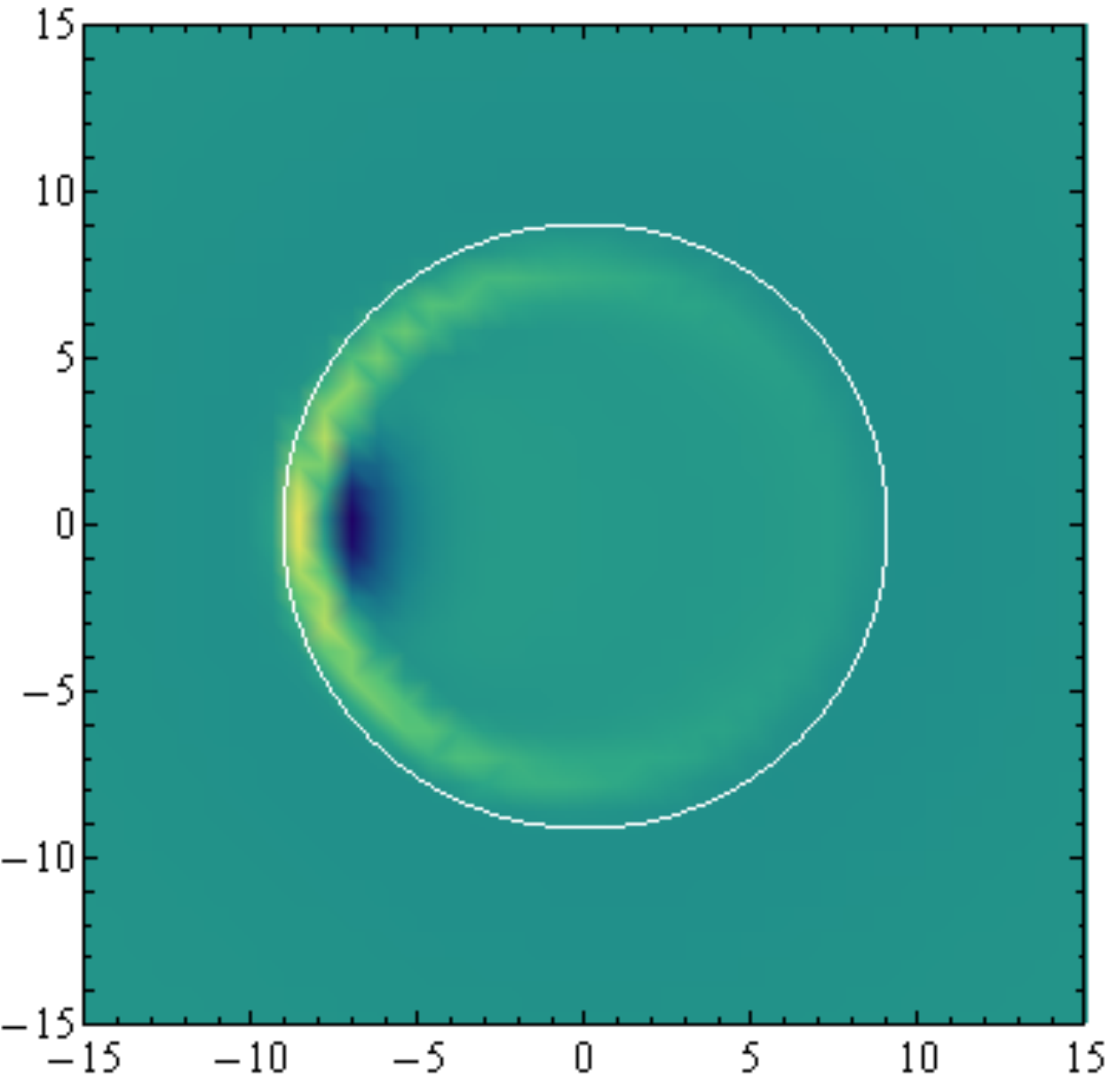}
  \caption{ Perturbation (arbitrary scale) of the temperature in the transverse plane (x,y) (in fm), induced by a jet
  generated at the point (0,0)  and stopped at (-8,0). The fireball rim is indicated by the wide circle.}
  \label{fig_U5}
\end{figure}

The second example, shown in  Fig.\ref{fig_U5}, is a jet originated at the fireball center and stopped near the rim.
It generates a large circle-shaped perturbation, but now the peak is in the  forward direction relative to the jet, or $\phi=\pi$.

Note, that in all these plots we had considered only one quenching jet, ignoring the signal from the 
trigger one. In this last case it will of course have the same travel path as the companion jet, and thus its quenching should be the same
but rotated by $\pi$ to $+x$ direction. So, the last
plot should be supplemented by its mirror image.

The next (and the last) example is the {\em asymmetric} jet, which has an impact parameter relative to the fireball center $b=3\, fm$. While the perturbation looks  again circle-like, its position and its maximum are
strongly displaced. 
The physical reason for that is ``cross wind" of the radial flow, indicated by thin arrows in Fig.\ref{fig_cone}.

The intersects of the perturbation with the   fireball rim (the white circle)  happen twice, the strongest at 
$\phi\approx 2.2$ and the weakest at $\phi\approx 1$. We thus conclude that such event would  generate two different-amplitude
peaks in the azimuthal distribution of associate particles.
Note that in this case  both peaks are located  rather far from the direction of the jet:
this happens due to the ``side wind" of the radial flow. Since their location depends strongly on
the particular geometry, after event averaging they all perhaps be averaged out.
Only in the case of very large energy deposition -- argued below to be perhaps observable on event-by-event 
bases -- one may have a chance to observed such strongly displaced jet shapes.

   \begin{figure}[t]
\includegraphics[width=8cm]{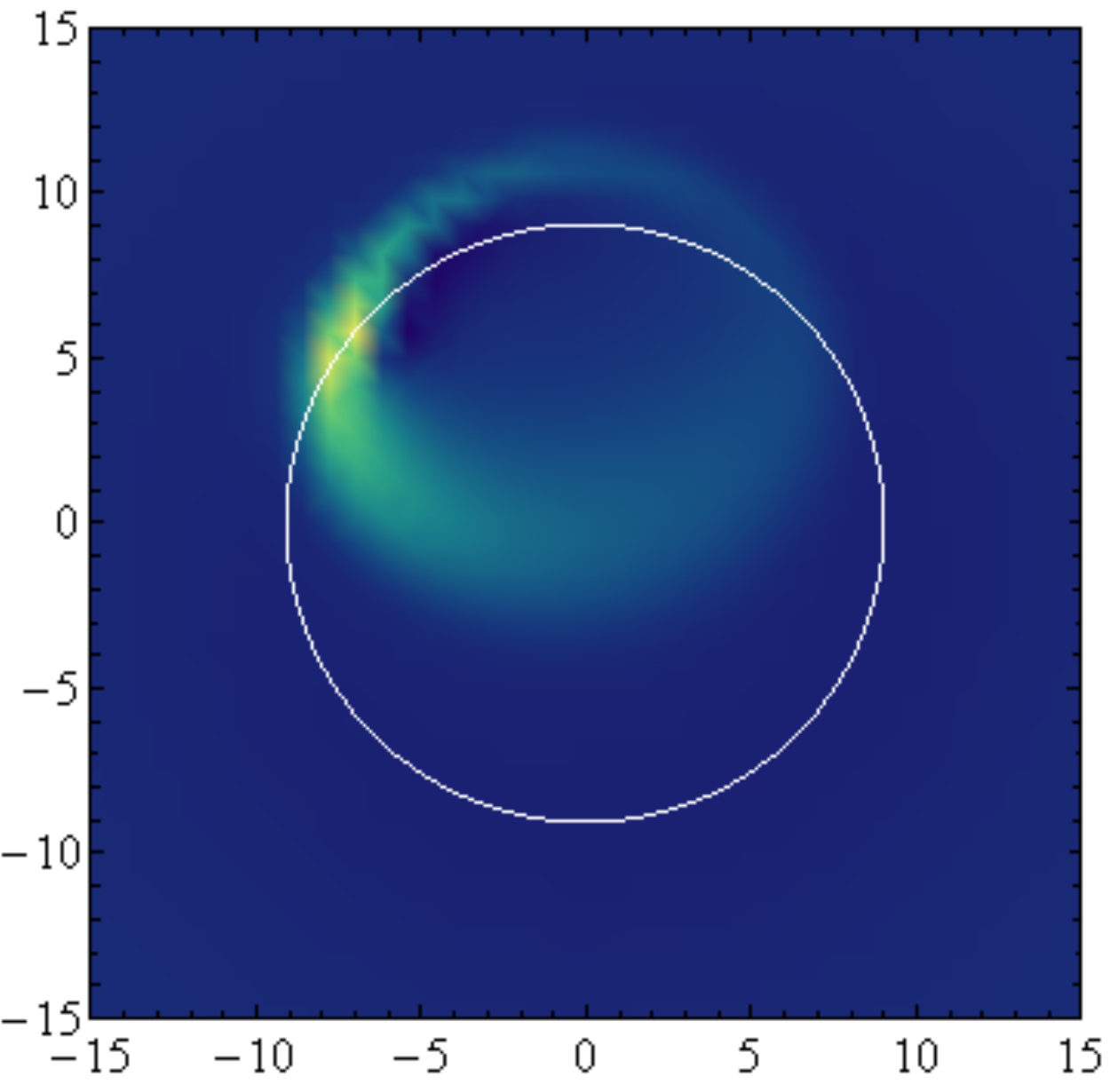}
  \caption{ Temperature perturbation (arbitrary scale) of the temperature in the transverse plane (x,y) (in fm), induced by a jet
  generated at the point (0,3) and stopped at the point (-5,3).  }
  \label{fig_U6}
\end{figure}

\subsection{Rapidity dependence} \label{sec_eta}
The calculations above were done in approximation including transverse coordinates but ignoring the longitudinal variable, the space-time rapidity. Now we restore summation over  
 the discretized momentum $k$ conjugated to $\eta$. 
Since $k$ appears in
the equations for the ``time" dependent function  $R_{l,k}(\rho)$
those are solved numerically for each $k,l$, 
 by Mathematica and/or fortran ODE solvers. The resulting multidimensional sums over $l,k$ and the energy deposition points along the jet line were done in Fortran.

\begin{figure}[!t]
\includegraphics[width=5cm]{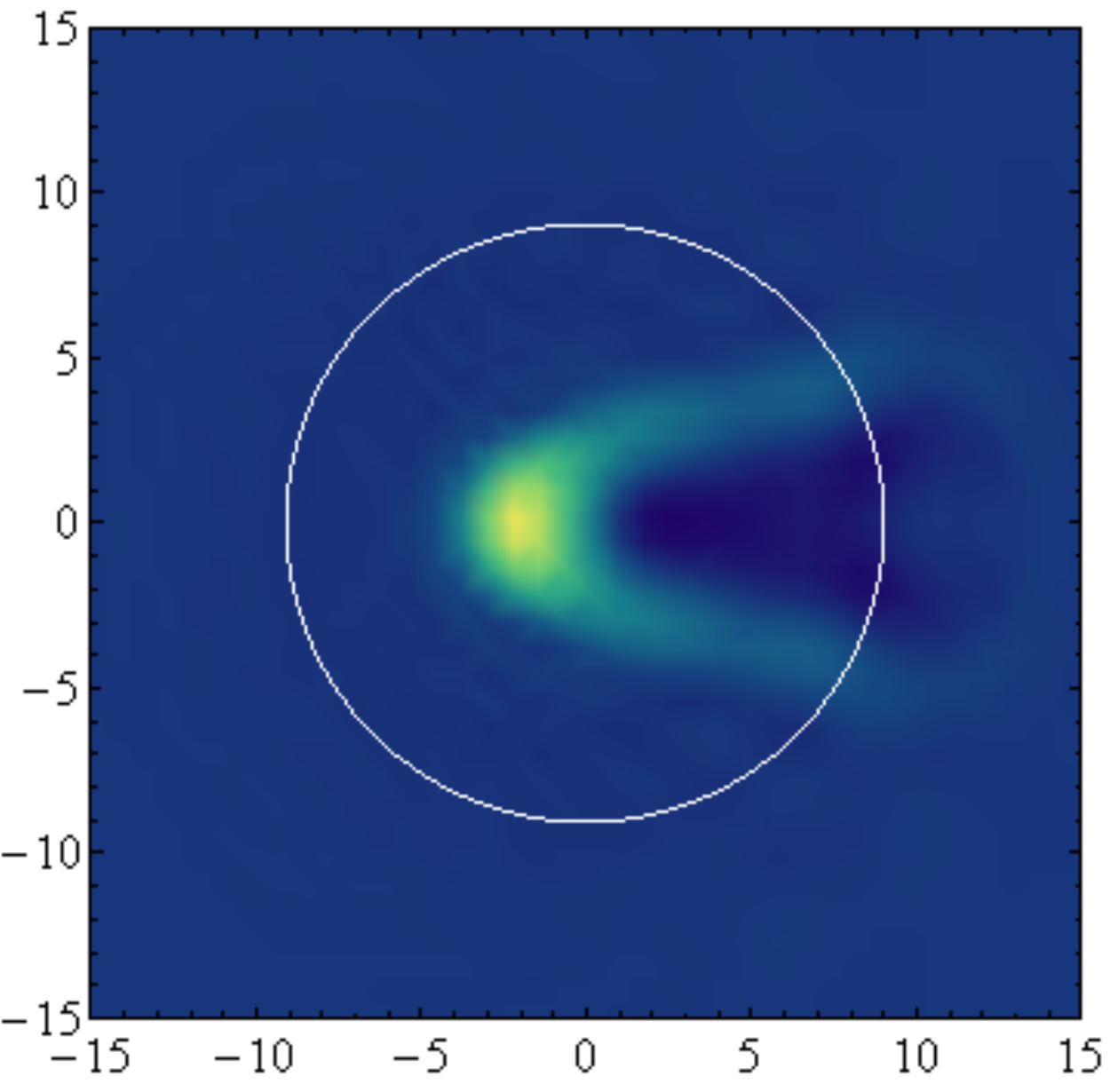}
\includegraphics[width=5cm]{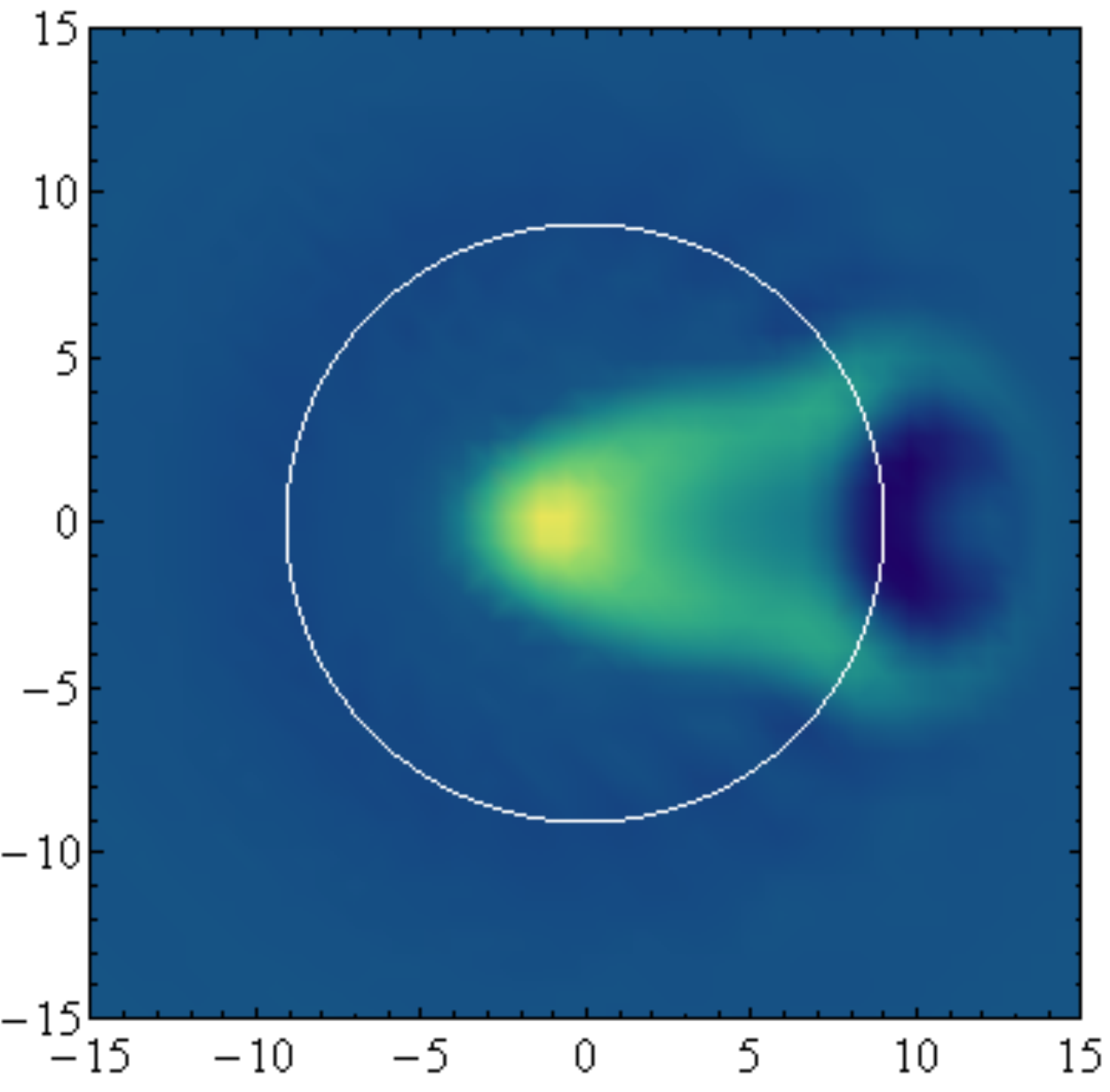}
\includegraphics[width=5cm]{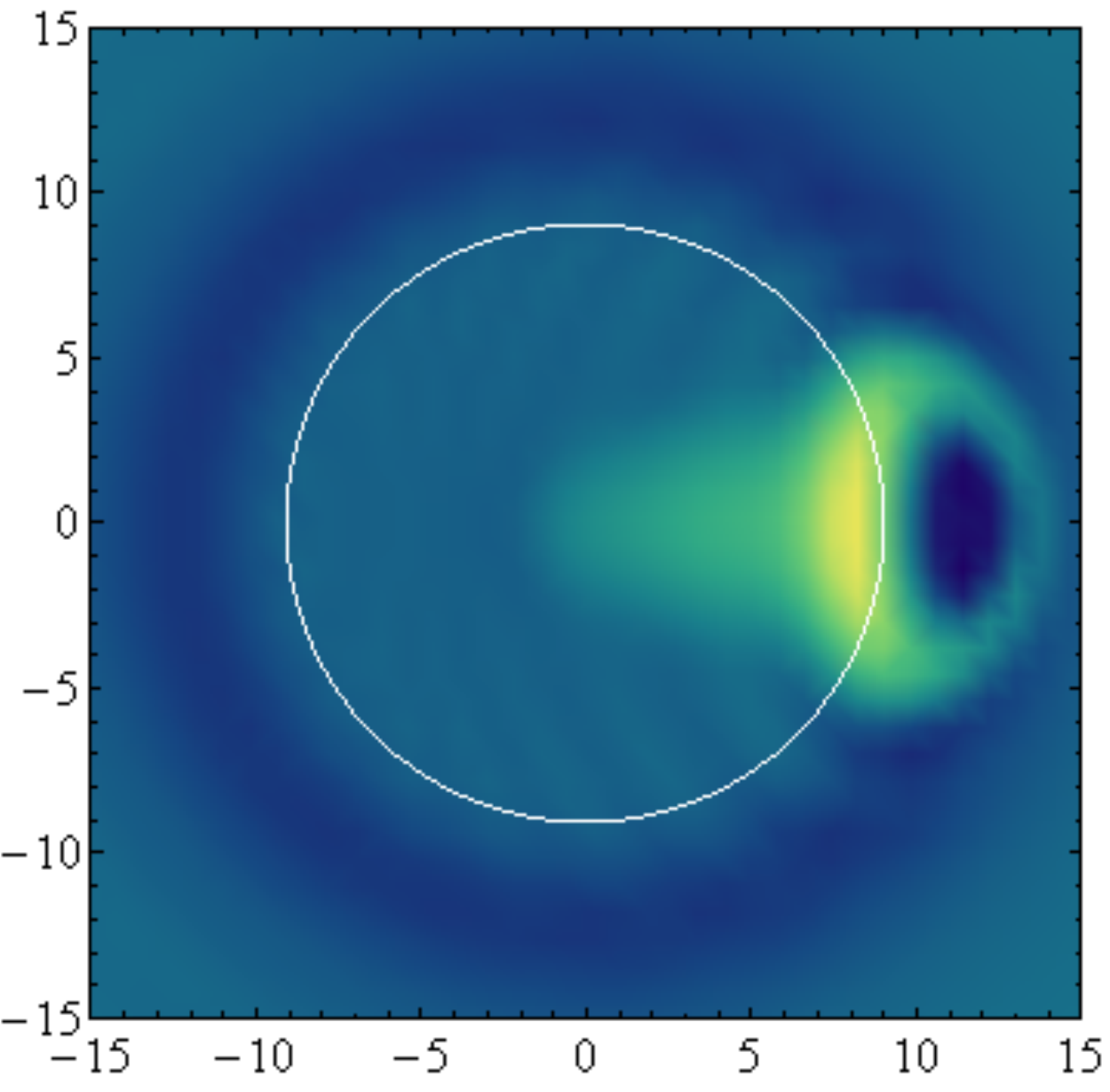}
\includegraphics[width=5cm]{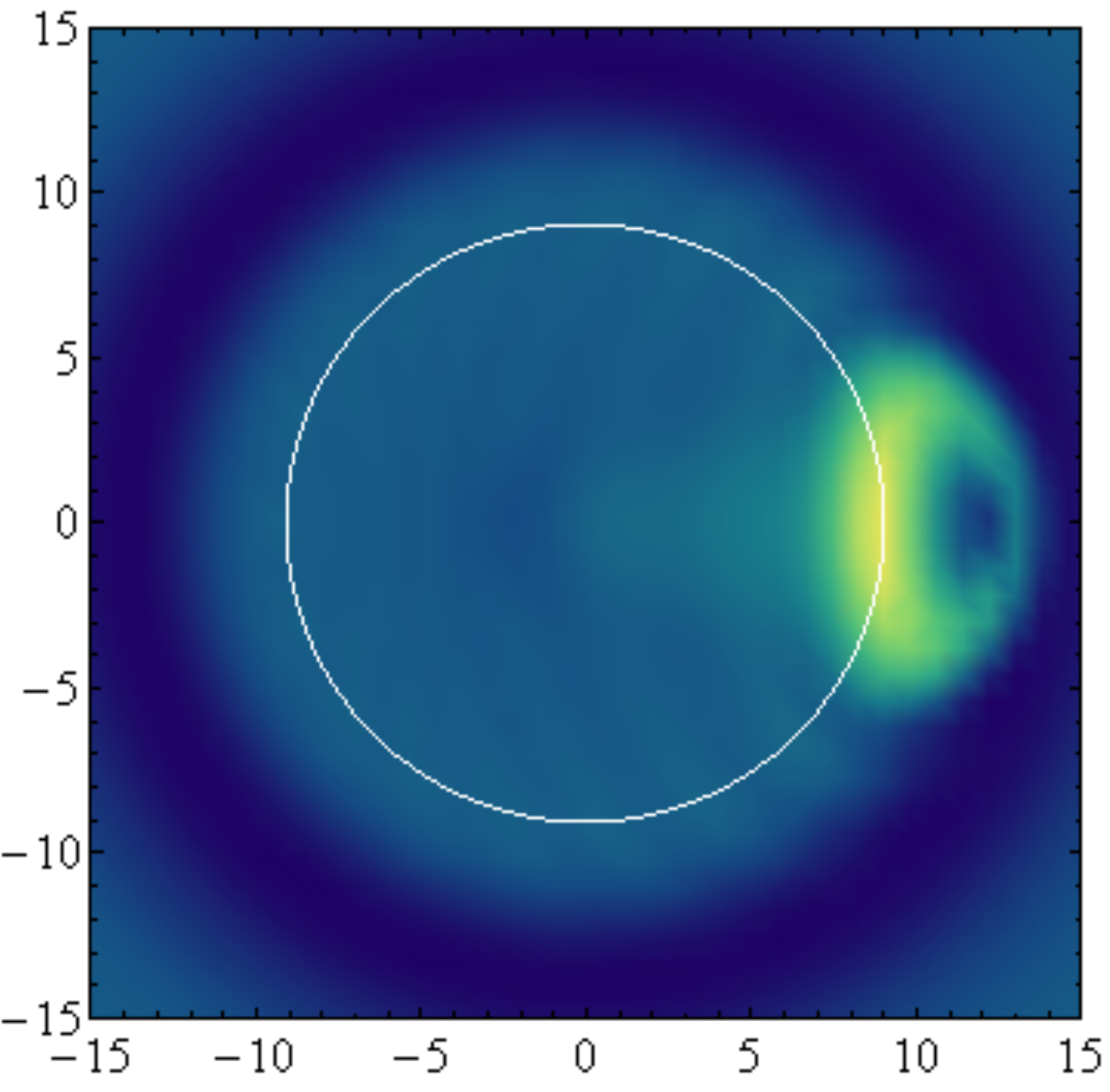}
  \caption{ Perturbation (arbitrary scale) of the temperature in the transverse plane (x,y) (in fm), induced by a jet
  generated at the point (6.1,0): the same case as in Fig.\ref{fig_U2} but with the $\eta$ variable included.
 Four pictures, top to bottom, are for $\eta=0,0.4,0.8$ and $1$ .  }
  \label{fig_eta1}
\end{figure}

In Fig.\ref{fig_eta1} we show the distribution, obtained in the same case as in Fig.\ref{fig_U2}, now in full 3+1 space-time
structure. The plots top-to-bottom show sections at variable eta. While the first plot at $\eta=0$ is hardly different from Fig.\ref{fig_U2} (a),
displaying the same Mach cone, the picture changes with increase $\eta$. Furthermore, at $\eta=0.8$ we see a 
quite brighter picture, indicated that here we hit the wall of the Mach cone
in the $\eta$ direction, which abruptly ends at the next picture at $\eta=1$.  If one projects it to the fireball edge (shown by the white circle)
one finds two peaks at certain $\pm \phi_{peak}$ increasing in amplitude till $\eta=0.8$, before it reduces to the single peak at $\phi=0,\eta>0.8$.

\begin{figure}[!t]
\includegraphics[width=5cm]{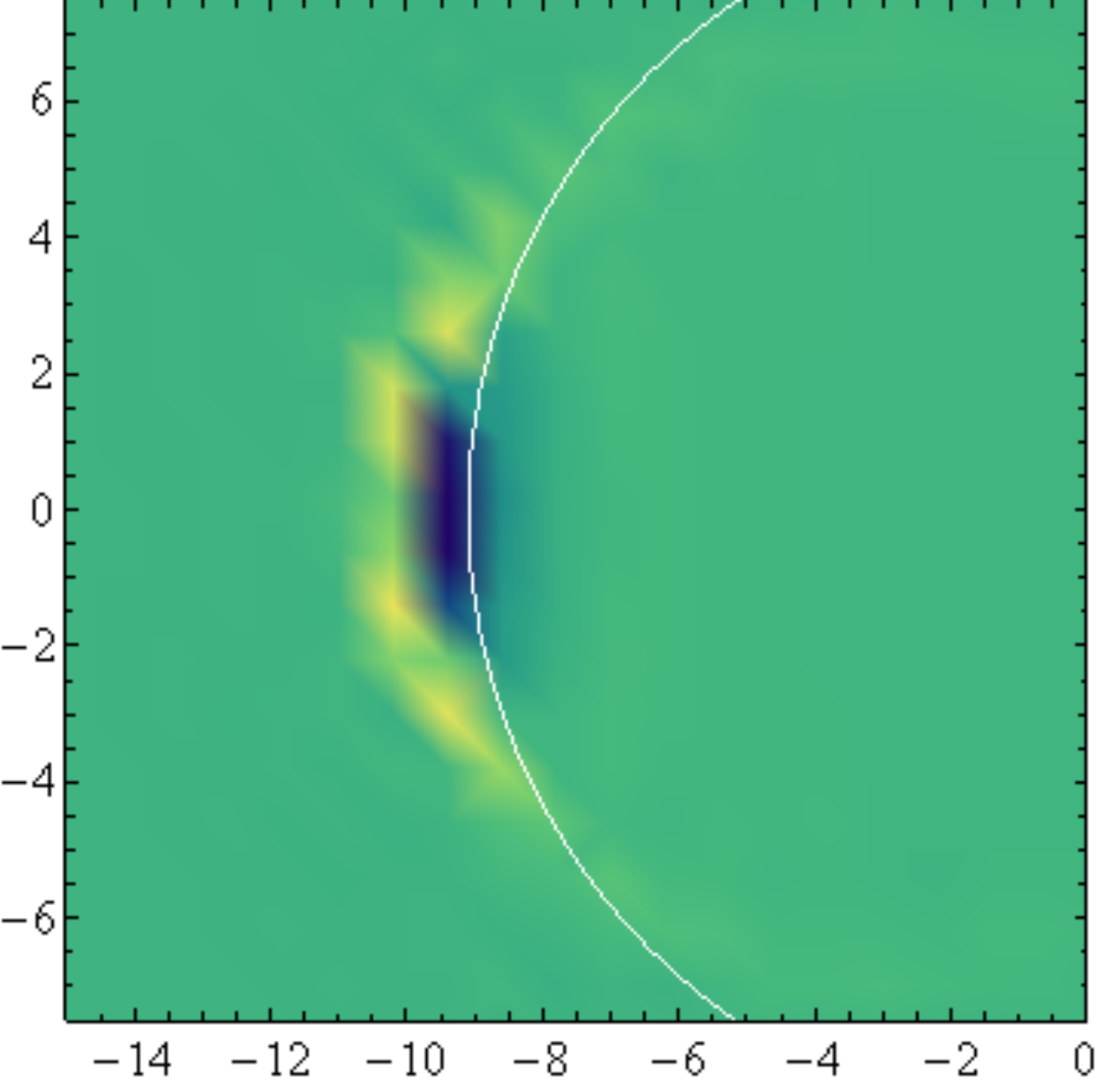}
\includegraphics[width=5cm]{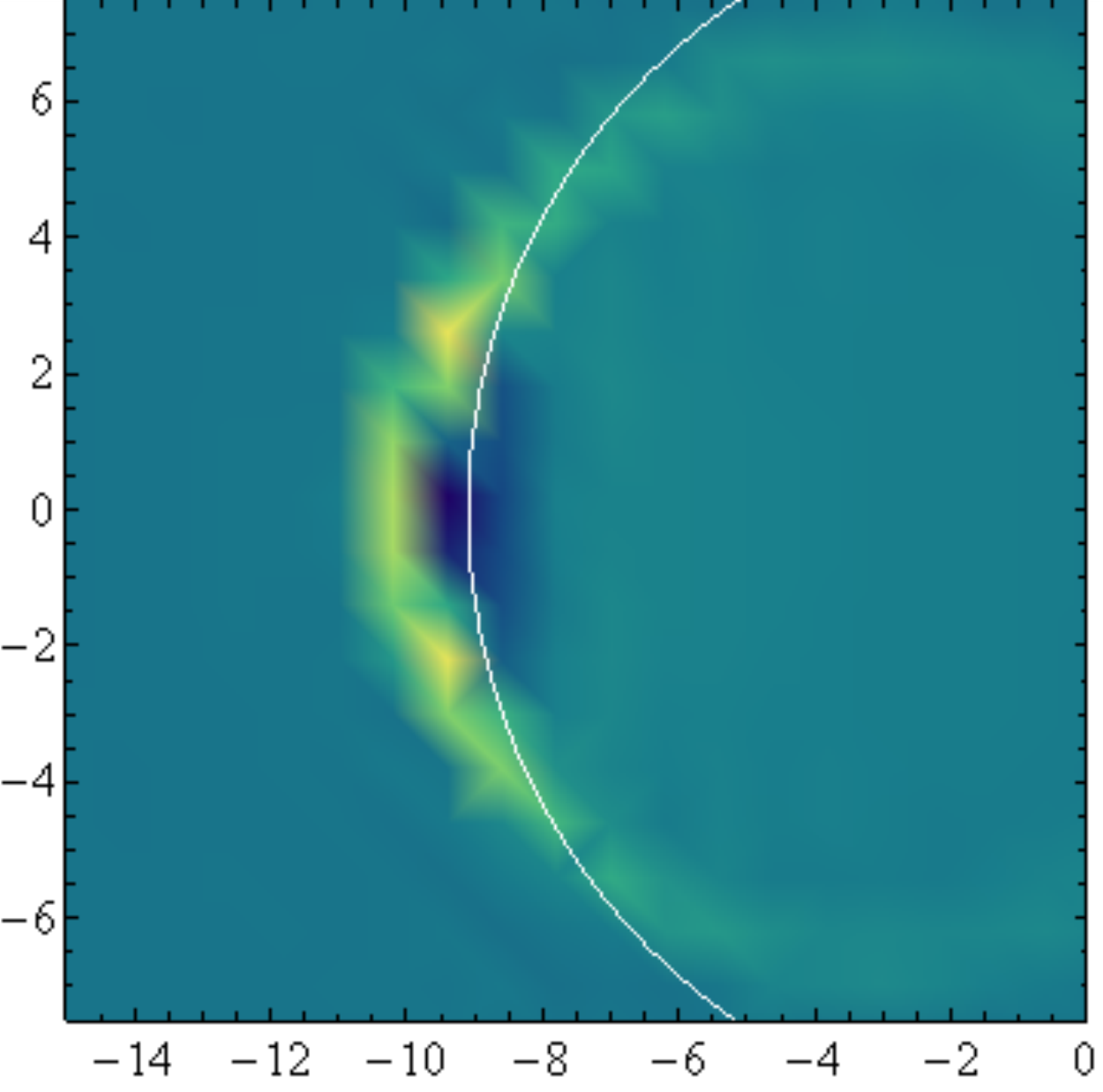}
\includegraphics[width=5cm]{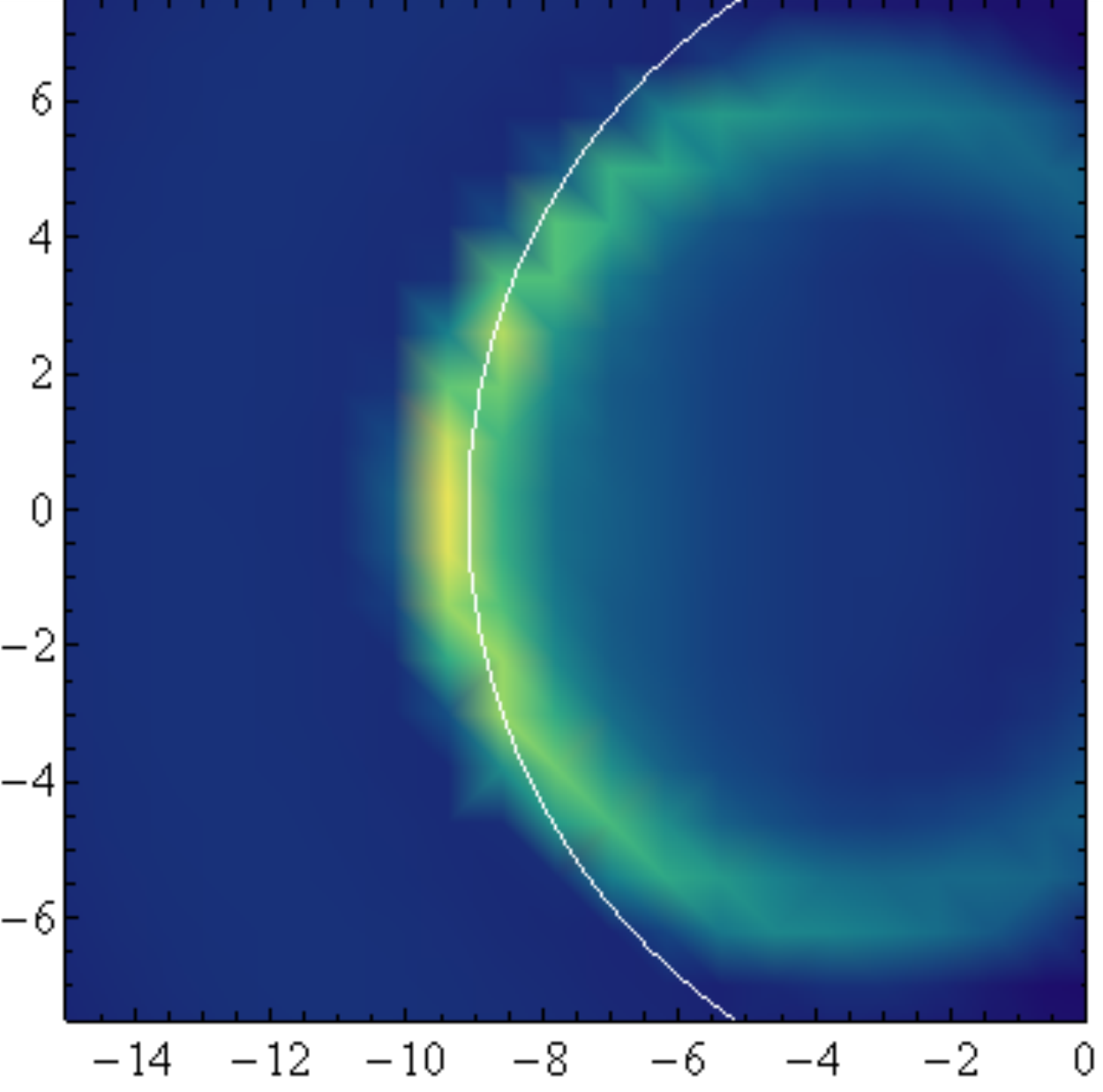}
\includegraphics[width=5cm]{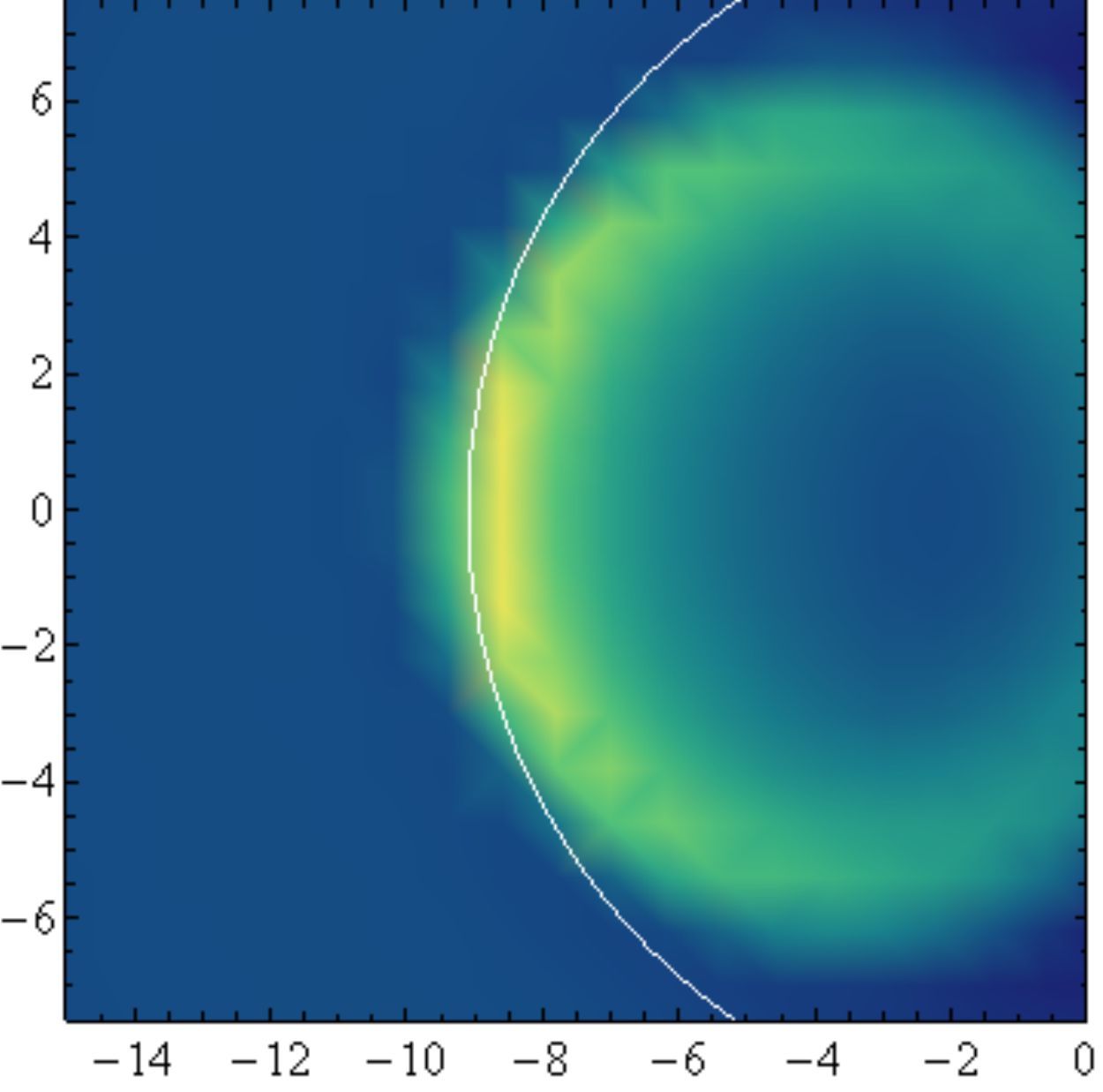}
  \caption{ Perturbation (arbitrary scale) of the temperature in the transverse plane (x,y) (in fm), induced by a jet
  generated at the point (-2,0) and stopped near the fireball edge.
 Four pictures, top to bottom, are for $\eta=0,0.4,0.8$ and $1$ .  }
  \label{fig_eta2}
\end{figure}

In Fig.\ref{fig_eta2} we show similarly located series of pictures, but now for the jet stopped near the fireball edge. What makes it different
is that this jet originates from the point (-2,0). As a consequence, 
between its stopping time and the freezeout there remains a certain time interval, allowing for the sound propagation from the last deposition point to proceed further. 

\begin{figure}[!t]
\includegraphics[width=6cm]{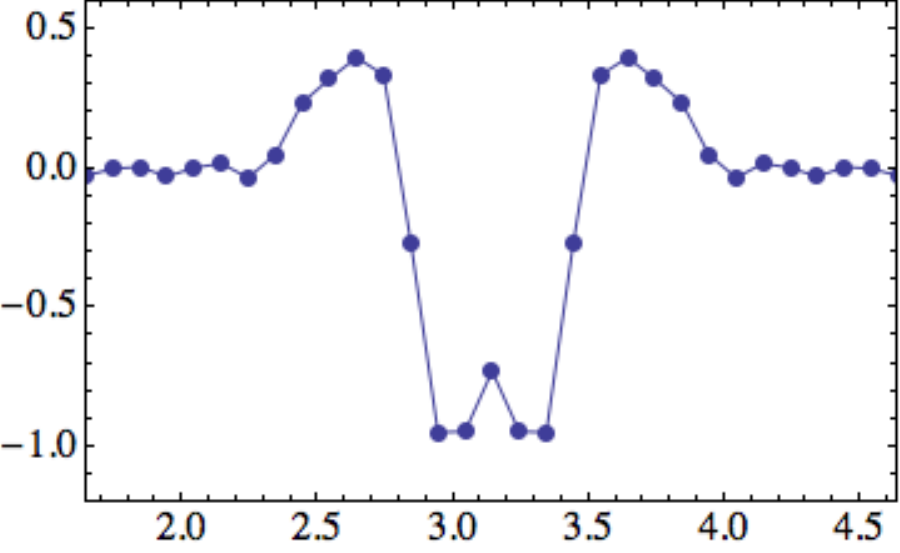}
\includegraphics[width=6cm]{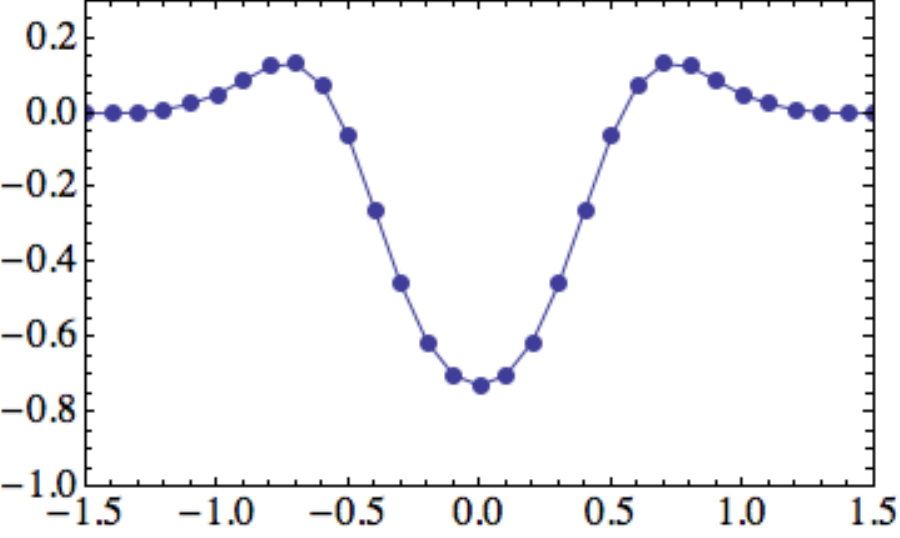}
  \caption{ Distribution over $\phi$ at $\eta=0$ (upper plot) and $\eta$ at $\phi=\pi$ (lower plot),
   projected on the fireball rim $r=9.1\, fm$.  The same case as in 
  the previous Fig.\ref{fig_eta2} .  }
  \label{fig_rim}
\end{figure}

The projection of the picture onto the fireball rim is shown in Fig.\ref{fig_rim}, now as a function of $\phi$ and $\eta$  separately.
Although the shapes in two direction is a bit different, the qualitative features are the same:
there is (i) a positive $\delta T$ at the  ``edge of the jet" (anticipated
in  Fig.\ref{fig_cone2}(b)), complemented by (ii) a deep depletion (negative $\delta T$) in the middle.
Altogether it shows a ``splash" of perturbed matter away from the jet.
The radius of the ``edge" depends on the time between the last energy deposition and freezeout.    
 
 (Let us at this point remind the reader, that
 according to kinematical arguments given above, a bit ``blurred" version of these
 pictures should  be translated into the corresponding momentum  variables   $\phi_p,y_p$ as well, since
 the Cooper-Fry integration is near-local. ) 

\section{Large energy deposition and bulk observables}   
   Examples given above support the idea, that $\delta T$ is mostly
 concentrated near the sound surface, with little effect in the sound-perturbed bulk.
   And yet, when one considers not the high-$p_\perp$ end of the spectra but bulk observables, 
   it may happen that large bulk volume  compensates for smallness of  $\delta T$ and the bulk contribution is important.


As seen from 
Fig.\ref{fig_cone2}(b), even the total volume of affected ``tube" of the medium is still a relatively small part of the total fireball. 
On the other hand, energy/momentum deposition into it can be substantial.   
     Let us provide a simple (upper limit) estimate of how different
this matter is from the unperturbed ambient matter. Using mid-rapidity (ALICE value)
of the  multiplicity $dN_{ch}/d\eta\approx 1584$ of the charged particle, we
multiply it by 3/2 to include neutrals and get $dN/d\eta\sim 2400$. Since the rapidity width of the region affected by a jet has $\Delta \eta \sim 1$, this multiplicity
can be directly compared with the ``extra particles" originated from the jet. At deposited $E_\perp\sim 100 \,GeV$ this number is about $N_{extra}\sim 200$,
provided they are fully equilibrated.  The multiplcity increase is about 8\%. Since multiplicity scales as $T^3$, the increase of the temperature
(if homogeneous) is about $\delta T/T\sim 2.7 \%$. 

At the freezeout the matter density is approximately constant, independent of collision energy. 
 (E.g. at twice larger multiplicity of LHC relative to RHIC one indeed finds twice large HBT volume, as shown by ALICE.)
  Therefore    $N_{extra}$ particles produced by jet energy deposition need about   8\% of extra freezeout volume. Assuming that longitudinal expansion is still rapidity-independent, it means increasing
the transverse area, or increasing the freezeout radius by the square root of it, or 4\% in our example. The Hubble law of expansion then tell us that it will increase flow
velocity linearly with $r$, or also by  4\%. The boost exponent however can easily increase the contrast  to be  as large as 100\%, for example
\be exp[({p_\perp u_t \over T_f}) {\delta u_t \over u_t} ]\sim exp(20*0.04)\sim 2.2  \ee
(using the same parameters  as in the example above).   
Such increase is  $O(1)$ in the kinematical window in question, and thus should be easily observable. 

Another effect stems from the fact that jets deposit not only energy but also equal amount of momentum, which does not
get lost by rescatterings.
 If e.g. it is fully equilibrated,  meaning that that matter inside the affected ``tube" gets extra directed flow velocity 
 \be \delta v= {P\over M} \sim 0.1 \ee
  where we used $P\sim 100\, GeV$ and the total mass of the affected matter $M\sim 1\, TeV$. This directed flow velocity is to be added to the extra radial flow estimated above. (Unlike enhanced radial flow directed radially outward, this
 $\delta v$ is directed along the jet.)

   In summary, in the events with very asymmetric dijets
  the ``underlying event" should
      be viewed as a superposition of two sub-events, with somewhat different properties.   While  outside of the sound surface 
there should be no   difference with the no-jet events,
      its inside should have noticiable extra radial and directed flow.
   
There are two standard strategies of observing flows. One is to measure 
the dependence of the mean $p_\perp$ (or the ``effective slopes" of the particle spectra) of secondaries on their mass. 
The other is to look into the kinematic window most affected by flow, $p_\perp =1- 3 \,GeV$, and simply 
 compare the number of  secondaries inside the 
  angular region     \be r_{jet}\approx 0.3 < \sqrt{ \Delta \phi^2 + \Delta y^2 } < r_{edge}\sim 1 \label{eqn_range} \ee
between the jet radius and the sound edge, with those in the unperturbed regions of the same events.
As all statements above, 
this one  is of course statistical, and should be studied for a sample of events with close energy deposition.

 One may further suggest that since the predicted contrast  can be large enough, one can  perhaps see its manifestation in
single events.  
In principle, this should happen both for companion and trigger jets, although with larger 
magnitude in the former case.


  \begin{figure}[t]
\includegraphics[width=8cm]{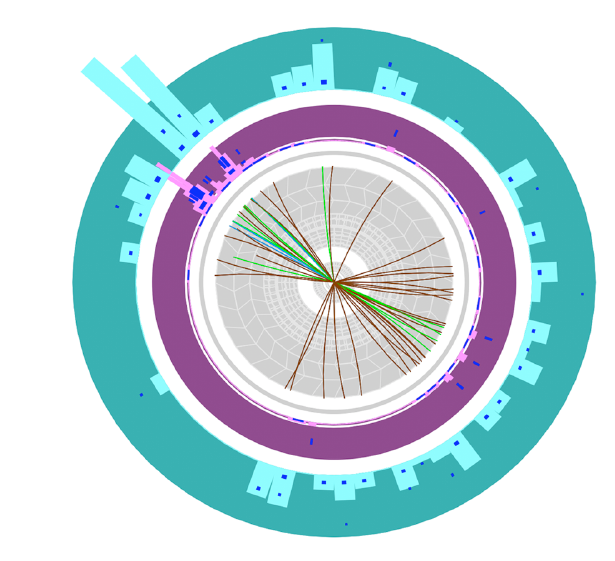}
  \caption{(color online) Azimuthal distribution of the transverse energy in one event from \cite{ATLAS_jets} ,  the inner part shows tracks with $p_\perp > 2.6 \, GeV$, the intermediate (pink)  histogram is the electromagnetic calorimeter energy with $E_t>0.7\, GeV$ threshold,
and the outer (blue) histogram is the  hadronic 
calorimeter energy distribution, with thresholds $E>1\, GeV$ per cell. 
 }
  \label{fig_ATLAS_event}
\end{figure}

 
One event display (from the  famous ATLAS  first paper on very asymmetric dijets at LHC  \cite{ATLAS_jets}) is reproduced in  our Fig. \ref{fig_ATLAS_event}; note that tracks $p_\perp$ and the calorimeter cuts crudely  correspond to the kinematic window we propose to be used for ``jet edge" observation. 
 The three histograms  are indeed very peculiar.
The charged tracks and the (outer) hadronic calorimeter  signal 
  are   not Gaussian-like but  have two regions with constant density, with identifiable edges, separating it from  two ``empty" regions at
 large angles $\phi_p=\pm \pi/2$  without any  signal. The (pink) signal from the EM calorimeter has two peaks around the jet axis.
  We suggest that those  distributions are related to  the  ``jet edge" phenomenon. 
 

\section{Summary and Discussion} 
   In this paper we studied the fate of the energy/momentum deposited by quenching jets into the medium.
   The technical method used has been a solution of (linearized) hydrodynamics, on top of (axially symmetric central)
``Gubser flow" solution.    
   
     On general grounds one expects a perturbation peaked at the Mach cone, supplemented by spheres
     around the origination and final points.  The explicit solution however had demonstrated that
the combined effect increasing energy deposition along the jet path  and viscosity substatially change the amplitude of the perturbation, placing more weight to later stages of the process. They significantly 
weaken the Mach cone part of the surface and enhance the role of  the last deposition point.  

    In a particular kinematical window $p_\perp \sim 2 \,GeV$ for the associate particle mostly the {\em intersection} of the perturbation and
    the fireball rim are contributing. Therefore, on the $\Delta \phi_p, \Delta y_p$ plane those should be seen as certain closed curves,
    or the ``sound circles".
    As we detailed above, for different geometries of the jet path those can have various locations. And yet  we think that
    the dominant case is the one in which they are close to the jet direciton   and form what we call ``the jet edge" 
    of elliptic shape, with  $\Delta \phi  \sim 1 \, rad, \Delta y\sim 1$. (The actual values depend
    on jet geometry, mostly on the time between the last deposition and the freezeout.)
If observed,  the ``jet edge" location can be used  as a tool to further constrain the geometry and the
     mechanism of jet quenching.
    
    We also point out that for dijet events with high energy/momentum deposition $\sim 100\, GeV$ the  part of the
      fireball inside the sound surface can be heated/boosted by an observable amount. If so, the whole underlying event should
      be viewed as a superposition of two sub-events, with somewhat different properties. 
      We propose in particular to compare the number of  secondaries with  $p_\perp =1- 3 \,GeV$ and their spectra in the 
   region    (\ref{eqn_range} ) with those in the regions far from the jet.
       Our estimates
show that       for this effect can become large enough to
be seen in individual effects.  If so, these observations can become even more valuable tool, providing
      information about geometry of the jet production on event-by-event basis.

     Going into discussion, we need certain number of disclaimers and calls for further scrutiny of the approximations used. Let us however mention only one example, related with the basic parameter,
 the speed of the perturbation. 
 
 We used the Gubser flow based on conformal QGP  with
the speed of sound $c_s=1/\sqrt{3}\approx 0.58$.  In matter undergoing QCD phase transition 
the speed of sound is not constant and is smaller, with a minimum near $T_c$.
    On the other hand, there is  an issue of finite amplitude perturbations.  As jet quenching results in local  energy deposition, 
    the perturbation of matter  in general starts as finite-amplitude one. If so,  disturbances will happen in the form of shocks, rather than sounds. 
    Unlike sounds, shocks  speed  depends on the perturbation amplitude. Those are larger than the speed of sound,
    only approaching them  with time as the shocks are weakening.
    Shocks in QGP were recently studied in  \cite{Shuryak:2012sf} and found to be produce relatively moderate corrections,
    as far as shock speed is concerned. For example,
    in the ideal QGP the shock with a factor of two jump in the pressure/energy density has the velocity $v_{shock}\approx 0.66$.
 Since those two effects act in the opposite directions, to some extent they cancel each other: thus  for simplicity we have ignored both and used a conformal QGP in this work. This approximation should obviously be studied and improved in subsequent works.

 {\bf Acknowledgments.} 
 supported in parts by the US-DOE grant
DE-FG-88ER40388.

\end{document}